\documentclass{inprep}
\usepackage{graphicx}

\newcommand{\al}{\alpha}
\newcommand{\be}{\beta}
\newcommand{\ga}{\gamma}
\newcommand{\de}{\delta}
\newcommand{\ep}{\varepsilon}
\newcommand{\te}{\vartheta}
\newcommand{\om}{\omega}
\newcommand{\si}{\sigma}
\renewcommand{\phi}{\varphi}

\newcommand{\Ga}{\Gamma}
\newcommand{\De}{\Delta}
\newcommand{\La}{\Lambda}
\newcommand{\Si}{\Sigma}
\newcommand{\Om}{\Omega}

\newcommand{\M}{{\cal M}}
\newcommand{\B}{{\cal B}}

\newcommand{\Q}{\widetilde{Q}}
\renewcommand{\S}{\widetilde{S}}
\newcommand{\Z}{\widetilde{Z}}
\renewcommand{\j}{\widetilde{\jmath}}
\newcommand{\J}{\widetilde{J}}
\newcommand{\p}{\widetilde{p}}
\newcommand{\f}{\widetilde{f}}
\newcommand{\e}{\widetilde{\varepsilon}}

\newcommand{\q}{\overline{q}}
\newcommand{\G}{\overline{\Gamma}}
\renewcommand{\u}{\overline{u}}
\newcommand{\m}{\overline{\mu}}
\newcommand{\MS}{$\overline{\rm MS}$}

\newcommand{\D}{\widehat{D}}
\renewcommand{\k}{\widehat{k}}
\renewcommand{\v}{\widehat{v}}

\newcommand{\va}[1]{{<}#1{>}}

\newcommand{\Tr}{\mathop{\rm Tr}\nolimits}
\newcommand{\ch}{\mathop{\rm ch}\nolimits}
\newcommand{\sh}{\mathop{\rm sh}\nolimits}
\renewcommand{\th}{\mathop{\rm th}\nolimits}
\newcommand{\cth}{\mathop{\rm cth}\nolimits}

\begin{document}

\begin{center}
\Large Budker Institute of Nuclear Physics
\\[35mm]
A.~G.~Grozin
\\[10mm]
Introduction to the Heavy Quark Effective Theory
\\[3mm]
\large Part 1
\\[20mm]
\Large BUDKERINP 92-97
\\
\vfill
NOVOSIBIRSK
\\[2mm]
1992
\end{center}
\thispagestyle{empty}
\newpage

\begin{center}
\bf
Introduction to the Heavy Quark Effective Theory
\\
\rm Part 1
\\[3mm]
\it A.~G.~Grozin$^{1,2}$
\\[3mm]
\rm Budker Institute of Nuclear Physics,
\\
630090 Novosibirsk 90, Russia
\\[5mm]
ABSTRACT
\end{center}
\begin{quotation}
Heavy Quark Effective Theory (HQET) is a new approach to QCD problems
involving a heavy quark. In the leading approximation, the heavy quark is
considered as a static source of the gluon field; $1/m$ corrections can be
systematically included in the perturbation theory. New symmetry properties
not apparent in QCD appear in HQET. They are used, in particular, to obtain
relations among heavy hadron form factors. HQET also simplifies lattice
simulation and sum rules analysis of heavy hadrons.

Part 1 contains discussion of the effective lagrangian, mesons, baryons,
and renormalization. Part 2 will contain $1/m$ corrections, nonleptonic
decays, and interaction with soft pions.
\end{quotation}
\vfill
\begin{flushright}
\copyright{} Budker Institute of Nuclear Physics
\end{flushright}
\addtocounter{footnote}{1}
\footnotetext{Supported in part by grant from the Soros fund}
\addtocounter{footnote}{1}
\footnotetext{Internet address: \tt GROZIN@INP.NSK.SU}
\thispagestyle{empty}
\newpage

{
\parindent=0pt
\begin{tabular}{r}
\hline
\hspace{112mm}
\end{tabular}
\par
}

\vspace{50mm}

\section{Effective Lagrangian}
\label{SecLagr} \setcounter{equation}{0}

Recently an interesting new approach to QCD problems involving a heavy quark
was proposed, namely the Heavy Quark Effective Theory (HQET). In the leading
approximation, the heavy quark is considered as a static source of the gluon
field; $1/m$ corrections can be systematically included in the perturbation
theory. This simplification is similar to considering a hydrogen atom instead
of a positronium. New symmetry properties not apparent in QCD appear in HQET.
They are used, in particular, to obtain relations among heavy hadron form
factors. However, in QCD even such a simplified problem is unsolvable.
Approximate methods such as lattice simulation or sum rules are necessary
to obtain quantitative results. Here again HQET allows to proceed much further
than QCD.

There are several good reviews of HQET~\cite{RevGeorgi,RevGrin,RevIW}
to which we address the reader for an additional information. Here we widely
use the properties of currents' correlators to obtain general results. This
approach is inspired by sum rules, though we shall not consider details of
sum rules calculations. We shall start from a very simple though approximate
picture in the Sections~\ref{SecLagr}--\ref{SecBaryon}; some complications
are discussed later.

Let's start from the QCD Lagrangian
\begin{equation}
L = \overline{Q}(i\D-m)Q + \q i\D q - \frac{1}{4}G^a_{\mu\nu}G^a_{\mu\nu}
  + \cdots
\label{QCDlagr}
\end{equation}
where $Q$ is the heavy quark field,
$q$ are light quark fields (their masses are not written down for simplicity),
$G^a_{\mu\nu}$ is the gluon field strength,
and dots mean gauge fixing and ghost terms.
It is well known that the free heavy quark Lagrangian
$\overline{Q}(i\widehat\partial-m)Q$
gives the dependence of the energy on the momentum
$\ep=\sqrt{m^2+\vec p\,^2}$. We shall consider problems with a single heavy
quark approximately at rest, and all characteristic momenta $|\vec p|\ll m$.
Then we can simplify the dispersion law to $\ep=m$. It corresponds to the
Lagrangian $\overline{Q}(i\ga_0\partial_0-m)Q$. In such problems it is
convenient to measure all energies relative to the level $m$. This means that
instead of the true energy $\ep$ we shall use the effective energy
$\e=\ep-m$. Then the heavy quark energy $\e=0$
independently on the momentum. The free Lagrangian giving such a dispersion
law is $\overline{Q}i\ga_0\partial_0 Q$. The spin of the heavy quark at rest
can be described by a 2-component spinor $\Q$ (we can also consider it as
a 4-component spinor with the vanishing lower components: $\ga_0\Q=\Q$).
Reintroducing the interaction with the gluon field by requirement of the gauge
invariance, we arrive at the HQET Lagrangian~\cite{EichtenHill}
\begin{equation}
L = \Q^+iD_0\Q + \q i\D q - \frac{1}{4}G^a_{\mu\nu}G^a_{\mu\nu}
  + \cdots
\label{HQETlagr}
\end{equation}
The static quark field $\Q$ contains only
annihilation operators. There are no heavy antiquarks in the theory,
because processes of heavy quark-antiquark pair production are suppressed
by $1/m$. The heavy antiquark (if present) is described by a separate
field. The field theory~(\ref{HQETlagr}) is not
Lorentz-invariant, because the heavy quark defines a selected frame---its
rest frame.

The Lagrangian~(\ref{HQETlagr}) gives the static quark propagator
\begin{equation}
\S(\p) = \frac{1}{\p_0+i0},\quad
\S(x) = \S(x_0)\de(\vec x),\quad
\S(t) = -i\te(t).
\label{HQETprop}
\end{equation}
In the momentum space it depends only on $\p_0$ but not on $\vec p$
because we have neglected the kinetic energy. Therefore in the coordinate
space the static quark does not move. The unit $2\times2$ matrix is assumed
in the propagator~(\ref{HQETprop}). It is often convenient
to use it as a $4\times4$ matrix; in such a case the projector
$\frac{1+\ga_0}{2}$ excluding the lower components is assumed. The static
quark interacts only with $A_0$; the vertex is $ig\de_{0\mu}t^a$.

One can watch how expressions for QCD diagrams tend to the corresponding HQET
expressions in the limit $m\to\infty$~\cite{Grinstein}. The QCD heavy quark
propagator is
\begin{equation}
S(p) = \frac{\hat p+m}{p^2-m^2}
     = \frac{m(1+\gamma_0)+\widehat{\p}}{2m\p_0+\p\,^2}
     = \frac{1+\gamma_0}{2\p_0}+{\rm O}\left(\frac{\p}{m}\right).
\label{QCDprop}
\end{equation}
A vertex $ig\ga_\mu t^a$ sandwiched between two projectors
$\frac{1+\ga_0}{2}$ may be replaced by $ig\de_{0\mu}t^a$ (one may insert 
the projectors at external heavy quark legs too).
Therefore any tree QCD diagram equals the
corresponding HQET one up to O$(\p/m)$ terms. In loops, momenta can be
arbitrarily large, and the relation~(\ref{QCDprop}) can break.
But regions of large loop momenta are excluded by the renormalization in both
theories, and for convergent integrals one may
use~(\ref{QCDprop}) (see Sec.~\ref{SecRenorm}).

The Lagrangian~(\ref{HQETlagr}) can be rewritten in covariant
notations:
\begin{equation}
L = \overline{\Q} iv_\mu D_\mu \Q + \cdots
\label{CovarLagr}
\end{equation}
where the static quark field $\Q$ is a 4-component spinor obeying the relation
$\v\Q=\Q$ and $v_\mu$ is the quark velocity. The true total momentum $p_\mu$
is related to the effective one $\p_\mu$ by
\begin{equation}
p_\mu=mv_\mu+\p_\mu, \quad |\p_\mu|\ll m.
\label{HQETp}
\end{equation}
The static quark propagator is
\begin{equation}
\S(\p) = \frac{1+\v}{2} \frac{1}{v_\mu\p_\mu+i0},
\label{CovarProp}
\end{equation}
and the vertex is $igv_\mu t^a$. In the limit $m\to\infty$ the heavy quark
can't change its velocity $v_\mu$ in any processes with bounded momenta
$\p_\mu$. Therefore there exists the velocity superselection rule%
~\cite{Georgi}: heavy quarks with each velocity $v_\mu$ can be treated
separately and described by a separate field $\Q_v$. If we are interested
in a transition of a heavy hadron with the velocity $v_1$ into a heavy
hadron with the velocity $v_2$, we can use the Lagrangian
\begin{equation}
L = \sum_i \overline{\Q}_i iv_{i\mu}D_\mu \Q_i + \cdots
\label{CovarLagr2}
\end{equation}
where $\Q_i$ is the static quark field with the velocity $v_i$ (the quark
$Q_1$ is present in the initial hadron and $Q_2$---in the final one). These
quarks have different propagators~(\ref{CovarProp}) and
vertices. They can be of the same or different flavour; it doesn't matter
because they can't transform into each other except by an external current
with an unbounded momentum transfer (of order $m$). It is even possible
to write a Lorentz-invariant Lagrangian~\cite{Georgi}
\begin{equation}
L = \int \frac{d^3\vec v}{2v_0} \overline{\Q}_v iv_\mu D_\mu \Q_v + \cdots
\label{GeorgiLagr}
\end{equation}
describing static quarks with all possible velocities at ones. But in any
specific problem only several heavy quarks with several velocities are
involved; all fields $\Q_v$ except few ones are in the vacuum state and
are irrelevant, and finite sums~(\ref{CovarLagr2}) are
sufficient.

There is an ambiguity what quark mass $m$ should be used in~(\ref{HQETp})%
~\cite{FalkNeubertLuke}. In general the HQET Lagrangian is
$\overline{\Q}(iv_\mu D_\mu-\de m)\Q$; the residual mass $\de m$ is shifted
when we change $m$. Physical quantities, of course, don't depend on this
choice. The most convenient definition of the heavy quark mass $m$ is one
that gives $\de m=0$. It corresponds to the pole of the quark propagator
at $v_\mu\p_\mu=0$, or $p^2=m^2$ in QCD. This pole mass is gauge-invariant.
There is also an ambiguity in the exact choice of $v_\mu$ in~(\ref{HQETp})%
~\cite{LukeManohar}. This reparametrization invariance relates coefficients
of terms of different orders in $1/m$ expansion. Quantization of the theory%
~(\ref{CovarLagr2}) was discussed in~\cite{DuganGoldGrin}.

The HQET Lagrangian~(\ref{HQETlagr}) possesses the $SU(2)$ spin
symmetry~\cite{IW}. The heavy quark spin does not interact with gluon field
in the limit $m\to\infty$ because its chromomagnetic moment vanishes. If
there are $n_h$ heavy quark flavours with the same velocity, there is the
$SU(2n_h)$ spin-flavour symmetry. For example, in the problem of transition
from a heavy hadron with the velocity $v_1$ to a different flavour heavy
hadron with the velocity $v_2$, at equal velocities $v_1=v_2$ the Lagrangian%
~(\ref{CovarLagr2}) has the $SU(4)$ spin-flavour symmetry which
relates all form factors to the form factor of a single hadron at zero
momentum transfer (equal 1). At non-equal velocities, it has only the
$SU(2)\times SU(2)$ spin symmetry relating form factors to each other.
The Lagrangian~(\ref{GeorgiLagr}) has the symmetry
$SU(2n_h)^\infty\times SO(3,1)$.

Not only the orientation but also the magnitude of the heavy quark spin is
irrelevant in HQET. This leads to a supersymmetry group called the
superflavour symmetry~\cite{Super}. It allows one to predict properties
of hadrons with a scalar or vector heavy quark appearing in supersymmetric
extensions of the Standard Model, in technicolor models, and in some composite
models. The scalar and vector static quark Lagrangians
\begin{equation}
L=\widetilde{\phi}^+ iD_0 \widetilde{\phi}+\cdots,\quad
L=\vec{\widetilde{V}}^+ iD_0 \vec{\widetilde{V}}+\cdots
\label{SuperLagr}
\end{equation}
have the $SU(n_h)$ and $SU(3n_h)$ spin-flavour symmetry. This idea can also
be applied to baryons with two heavy quarks~\cite{SavageWise} because they form
a small size (of order $1/m\al_s$) spin 0 or 1 bound state antitriplet
in color.

HQET has great advantages over QCD in lattice simulation of heavy quark
problems. Indeed, the applicability conditions of the lattice approximation
to problems with light hadrons are that the lattice spacing is much less than
the characteristic hadron size, and the total lattice length is much larger
than this size. For simulation of QCD with a heavy quark, the lattice
spacing must be much less than the heavy quark Compton wavelength $1/m$.
For $b$ quark it is impossible at present. The HQET Lagrangian does not
involve the heavy quark mass $m$, and the applicability conditions of the
lattice approximation are the same as for light hadrons~\cite{Eichten}.
Relation of the lattice HQET to the continuum one was investigated in~%
\cite{Boucaud,EichtenHill2,Maiani}.
Simulation results can be found in~\cite{Sim}.

\section{Mesons}
\label{SecMeson} \setcounter{equation}{0}

Due to the heavy quark spin symmetry, hadrons may be classified according to
the light fields' angular momentum and parity $j^P$ which are conserved
quantum numbers. In other words, we can switch off the heavy quark spin
using the superflavour symmetry, and then the hadron's momentum and parity
will be $j^P$. The $\overline{Q}q$ mesons are the QCD analog of the hydrogen
atom. The ground-state ($S$-wave) meson has $j^P=\frac12^+$; the excited
$P$-wave mesons have $j^P=\frac12^-$ and $\frac32^-$. When we switch the heavy
quark spin on, each of these mesons becomes a degenerate doublet. Its
components are transformed into each other by the heavy quark spin symmetry
operations. We have the ground-state doublet $0^-$, $1^-$, and the excited
$P$-wave doublets $0^+$, $1^+$, and $1^+$, $2^+$. Splittings in these doublets
(hyperfine splittings) are due to the heavy quark chromomagnetic moment
interaction violating the spin symmetry, and are proportional to $1/m$
(Sect.~5). 

\begin{sloppypar}
Form factors of the ground-state mesons in HQET were considered
in~\cite{IW,FGGW,Hussain}; applications to semileptonic $B$ decays
were discussed in the review~\cite{RevNeubert} and papers~\cite{Vcb},
and to $e^+e^-$ annihilation---in~\cite{e+e-1,e+e-2}.
Transition form factors to the $P$-wave mesons were considered
in~\cite{IW2,Balk}. A general method of counting independent form factors
applicable both to mesons and baryons was proposed in~\cite{Politzer},
and an elegant explicit construction---in~\cite{Falk}. It was applied to
ground state to arbitrary excited meson transitions in~\cite{HighMeson}.
Two-point HQET sum rules were investigated in~\cite{Shuryak,HQETsr2}, and
three-point ones---in~\cite{HQETsr3}. Mesons in two-dimensional QCD with
the large number of colors were considered in~\cite{Dim2}. Here we shall use
correlators of currents with the quantum numbers of mesons in order
to investigate properties of mesons in HQET.
\end{sloppypar}

When the heavy quark is scalar, there is one bilinear heavy-light current
without derivatives $\j_s=\Q_s^+q$ ($\Q_s^+$ is the heavy antiquark field).
It has no definite parity; the currents $\j_\pm=\frac{1\pm\ga_0}{2}\j_s$
have the parity $P=\pm1$ because the $P$-conjugation acts as $q\to\ga_0q$.
The current $\j_+$ has the quantum numbers of the ground-state $\frac12^+$
meson, and $\j_-$---of the $P$-wave $\frac12^-$ meson. Currents with the
quantum numbers of mesons with higher $j$ necessarily involve derivatives.

In the case of real-world spin $\frac12$ heavy quark, there are 4 bilinear
currents without derivatives $\j=\overline{\Q}\Ga q$. Indeed,
because of $\ga_0\Q=\Q$, the current with $\Ga=\ga_0$ reduces to $\Ga=1$;
$\ga_0\ga_5$---to $\ga_5$; $\si_{0i}$---to $\ga_i$; $\si_{ij}$---to
$\ep_{ijk}\ga_k\ga_5$. We are left with $\Ga=\ga_5$, $\vec\ga$ and $1$,
$\vec\ga\ga_5$. The first pair with $\Ga$ anticommuting with $\ga_0$
has the quantum numbers of the ground-state $0^-$, $1^-$ doublet; the second
pair with $\Ga$ commuting with $\ga_0$---of the $P$-wave $0^+$, $1^+$ doublet.

\begin{figure}[h]
\begin{center}
\includegraphics{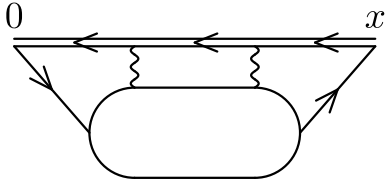}
\end{center}
\caption{Correlator of two HQET heavy-light currents}
\label{FigCorr2}
\end{figure}

A correlator of any two currents containing the static quark field has the
form (Fig.~\ref{FigCorr2})
\begin{eqnarray}
&&i\va{T\j_2(x)\j_1^+(0)}=\de(\vec x)\Pi(x_0),
\label{Corr2}\\
&&\Pi(\om)=\int\Pi(t)e^{i\om t}dt,
\quad
\Pi(t)=\int\Pi(\om)e^{-i\om t}\frac{d\om}{2\pi}.
\nonumber
\end{eqnarray}
It obeys the dispersion representation
\begin{equation}
\Pi(\om)=\int\limits_0^\infty\frac{\rho(\ep)d\ep}{\ep-\om-i0}+\cdots,
\quad
\Pi(t)=-\S(t)\int\limits_0^\infty\rho(\om)e^{-i\om t}d\om+\cdots
\label{Disp2}
\end{equation}
A subtraction polynomial in $\Pi(\om)$ (denoted by dots) gives $\de(t)$ and
its derivatives in $\Pi(t)$. We can analytically continue a correlator from
the half-axis $t>0$ to imaginary $t=-i\tau$. Then $\Pi(\tau)$ and $\rho(\om)$
are related by the Laplace transform
\begin{equation}
\Pi(\tau)=i\int\limits_0^\infty\rho(\om)e^{-\om\tau}d\om,
\quad
\rho(\om)=\frac1{2\pi}\int\limits_{a-i\infty}^{a+i\infty}
\Pi(\tau)e^{\om\tau}d\tau,
\label{Laplace2}
\end{equation}
where $a$ is to the right from all singularities of $\Pi(\tau)$.

The contribution of an intermediate state $|h{>}$ 
with the energy $\e$ to $\Pi(t)$, $\Pi(\om)$, $\rho(\om)$ is
\begin{eqnarray}
\Pi_h(t)&=&i\va{0|\j_2|h}i\S(t)e^{-i\e t}\va{h|\j_1^+|0},
\nonumber\\
\Pi_h(\om)&=&i\va{0|\j_2|h}\frac{i}{\om-\e+i0}\va{h|\j_1^+|0},
\nonumber\\
\rho_h(\om)&=&\va{0|\j_2|h}\va{h|\j_1^+|0}\de(\om-\e).
\label{Hadr2}
\end{eqnarray}
We  remind the reader that the HQET energy $\e$ means the true energy minus
the heavy quark mass.

The correlator of two meson currents with the scalar heavy quark has the
$\ga$-matrix structure (Fig.~\ref{FigCorr2})
\begin{equation}
i\va{T\j_s(x)\overline{\j}_s(0)}=\de(\vec x)\Pi_s(x_0),
\quad
\Pi_s=A+B\ga_0.
\label{Scal2}
\end{equation}
For the currents with the definite parity $P$ we have
\begin{equation}
i\va{T\j_P(x)\overline{\j}_P(0)}=\de(\vec x)\Pi_P(x_0)\frac{1+P\ga_0}{2},
\Pi_P=A+PB=\frac14\Tr(1+P\ga_0)\Pi_s.
\label{ScalP2}
\end{equation}
Due to the linear relations~(\ref{Corr2}--\ref{Laplace2}),
the same $\ga$-matrix structures and relations between $\Pi_s$ and $\Pi_P$
hold in both the coordinate space and the momentum one, and also for
spectral densities. When calculating the correlator using the Operator Product
Expansion (OPE), even-dimensional terms contain an odd number of $\ga$-%
matrices along the light quark line and after all integrations contribute
to $B$; odd-dimensional terms contain even number of $\ga$-matrices and
contribute to $A$. If we denote $\va{0|\j_+|M,\frac12^+}=\f_{M,\frac12^+}u$
where $u$ is the meson $M$ wave function, then the meson's contribution to
$\rho_s(\om)$ summed over polarizations is
$\f_{M,\frac12^+}^2\de(\om-\e)\sum u\overline{u}$, or the contribution to
$\rho_+(\om)$ is $\f_{M,\frac12^+}^2\de(\om-\e)$. Similar formulae hold
for $\frac12^-$ mesons.

Now let's switch the heavy quark spin on. The correlator is
(Fig.~\ref{FigCorr2})
\begin{equation}
i\va{T\j_2(x)\j_1^+(0)}=\de(\vec x)\Pi(x_0),
\quad
\Pi=\Tr\Ga_2\frac{1-\ga_0}{2}\overline{\Ga}_1\Pi_s.
\label{Real2}
\end{equation}
In $\Pi_s$, $\ga_0$ may be replaced by $P=\pm1$ for $\Ga_{1,2}$ (anti-)
commuting with $\ga_0$, and $\Pi_s$ becomes the scalar function $\Pi_P$:
\begin{equation}
\Pi=\Pi_P\Tr\Ga_2\frac{1-\ga_0}{2}\overline{\Ga}_1.
\label{RealP2}
\end{equation}
The correlators of the currents $\overline{\Q}\ga_5 q$ and
$\overline{\Q}\ga_i q$ with the quantum numbers of the ground-state $0^-$,
$1^-$ mesons are equal to $2\Pi_+$ and $2\de_{ij}\Pi_+$. If we denote
$\va{0|\overline{\Q}\ga_5 q|M,0^-}=\f_{M,0^-}$,
$\va{0|\overline{\Q}\vec\ga q|M,1^-}=\f_{M,1^-}\vec e$ (where $\vec e$
is the $1^-$ meson's polarization vector) then the meson's contributions
to the spectral densities are $\f_{M,0^-}^2\de(\om-\e_{0^-})$ and
$\de_{ij}\f_{M,1^-}^2\de(\om-\e_{1^-})$. Therefore the spin symmetry requires
that the mesons in $0^-$ and $1^-$ channels are degenerate 
($\e_{0^-}=\e_{1^-}=\e$), and
$\f_{M,0^-}=\f_{M,1^-}=\sqrt{2}\f_{M,\frac12^+}$. Similar formulae hold
for $P$-wave $0^+$, $1^+$ mesons.

\begin{sloppypar}
In QCD, the meson constants are usually defined as
$\va{0|\overline{Q}\gamma_\mu\gamma_5 q|M,0^-}=f_{M,0^-}p_\mu$,
$\va{0|\overline{Q}\gamma_\mu q|M,1^-}=m f_{M,1^-}e_\mu$,
where the meson states are normalized in the relativistic way:
$\va{M,\vec p\,'|M,\vec p}=2p_0\de(\vec p\,'-\vec p)$. This normalization
is senseless in HQET; we must use the non-relativistic normalization
$\va{M,\vec p\,'|M,\vec p}=\de(\vec p\,'-\vec p)$ instead. Then the
definitions read
$\sqrt{2m}\va{0|\overline{Q}\ga_0\ga_5 q|M,0^-}=m f_{M,0^-}$,
$\sqrt{2m}\va{0|\overline{Q}\vec\ga q|M,1^-}=m f_{M,1^-}\vec e$.
Finally we obtain the scaling law
\end{sloppypar}
\begin{equation}
f_{M,0^-}=f_{M,1^-}=\frac{2\f_{M,\frac12^+}}{\sqrt{m}}.
\label{Scaling}
\end{equation}

\begin{figure}[h]
\begin{center}
\includegraphics{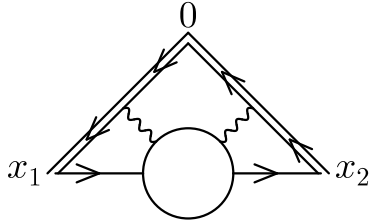}
\end{center}
\caption{Correlator of two HQET heavy-light currents and a heavy-heavy current}
\label{FigCorr3}
\end{figure}

To investigate hadron form factors in HQET, we consider correlators of two
currents $\j_{1,2}$ containing the static quark fields $\Q_{1,2}$ with the
velocities $v_{1,2}$ and the heavy-heavy velocity-changing current $\J$
(Fig.~\ref{FigCorr3}):
\begin{eqnarray}
&&\hspace{-4pt}
i^2\va{T\j_2(x_2)\J(0)\j_1^+(x_1)}=
\int\limits_0^\infty dt_2\de(x_2-v_2 t_2)
\int\limits_0^\infty dt_1\de(x_1+v_1 t_1) K(t_2,t_1),
\nonumber\\&&\hspace{-4pt}
K(\om_2,\om_1)=\int K(t_2,t_1)e^{i\om_2 t_2+i\om_1 t_1}dt_2 dt_1,
\label{Corr3}\\&&\hspace{-4pt}
K(t_2,t_1)=\int K(\om_2,\om_1)e^{-i\om_2 t_2-i\om_1 t_1}
\frac{d\om_2}{2\pi} \frac{d\om_1}{2\pi}.
\nonumber
\end{eqnarray}
They obey the double dispersion representation
\begin{eqnarray}
&&K(\om_2,\om_1)=\int\frac{\rho(\ep_2,\ep_1)d\ep_2 d\ep_1}
{(\ep_2-\om_2-i0)(\ep_1-\om_1-i0)}+\cdots
\label{Disp3}\\
&&K(t_2,t_1)=\S(t_2)\S(t_1)\int\rho(\om_2,\om_1)e^{-i\om_2 t_2-i\om_1 t_1}
d\om_2 d\om_1+\cdots
\nonumber
\end{eqnarray}
Subtraction terms in $K(\om_2,\om_1)$ (denoted by dots) are polynomial in
$\om_1$ with coefficients that are arbitrary functions of $\om_2$ (given
by single dispersion integrals) plus vice versa. These terms give $\de(t_1)$
and its derivatives times arbitrary functions of $t_2$ plus vice versa in
$K(t_2,t_1)$. We can analytically continue $K(t_2,t_1)$ from $t_{1,2}>0$
to $t_{1,2}=-i\tau_{1,2}$. Then $K(\tau_2,\tau_1)$ and $\rho(\om_2,\om_1)$
are related by the double Laplace transform
\begin{eqnarray}
&&K(\tau_2,\tau_1)=-\int\rho(\om_2,\om_1)e^{-\om_2\tau_2-\om_1\tau_1}
d\om_2 d\om_1,
\label{Laplace3}\\
&&\rho(\om_2,\om_1)=\frac1{(2\pi)^2}\int\limits_{a-i\infty}^{a+i\infty}d\tau_2
\int\limits_{a-i\infty}^{a+i\infty}d\tau_1 K(\tau_2,\tau_1)
e^{\om_2\tau_2+\om_1\tau_1}.
\nonumber
\end{eqnarray}

\begin{sloppypar}
The contribution of intermediate states $|h_{1,2}{>}$ 
with the energies $\e_{1,2}$ to $K(\tau_2,\tau_1)$, $K(\om_2,\om_1)$, and
$\rho(\om_2,\om_1)$ is
\begin{eqnarray}
&&\hspace{-20pt}
K_{h_2 h_1}(t_2,t_1)=i^2\va{0|\j_2|h_2}i\S(t_2)e^{-i\e_2 t_2}
\va{h_2|\J|h_1} i\S(t_1)e^{-i\e_1 t_1} \va{h_1|\j_1^+|0},
\nonumber\\&&\hspace{-20pt}
K_{h_2 h_1}(\om_2,\om_1)=i^2\va{0|\j_2|h_2}\frac{i}{\om_2-\e_2+i0}
\va{h_2|\J|h_1} \frac{i}{\om_1-\e_1+i0} \va{h_1|\j_1^+|0},
\nonumber\\&&\hspace{-20pt}
\rho_{h_2 h_1}(\om_2,\om_1)=\va{0|\j_2|h_2}\va{h_2|\J|h_1}\va{h_1|\j_1^+|0}
\de(\om_2-\e_2)\de(\om_1-\e_1),
\label{Hadr3}
\end{eqnarray}
where the sum over $h_{1,2}$ polarizations is assumed. Let's introduce the
spin wave functions $\psi_{1,2}$ of $h_{1,2}$. Then
$\va{0|\j_{1,2}|h_{1,2}}=\f_{1,2}\psi_{1,2}$,
$\va{h_2|\J|h_1}=\psi_2^+\f_{21}\psi_1$, where $\f_{21}$ is a form factor
matrix in the spin space. The spectral density of the correlator of the
currents $\j_{1,2}^+\psi_{1,2}$ with some specific polarizations $\psi_{1,2}$
is $\psi_2^+\rho(\om_2,\om_1)\psi_1$. The contribution of $|h_{1,2}{>}$
to it is $\f_2\f_1\psi_2^+\f_{21}\psi_1\de(\om_2-\e_2)\de(\om_1-\e_1)$.
\end{sloppypar}

The correlator of two meson currents with the scalar heavy quark $\j_{s2}$,
$\overline{\j}_{s1}$ and the scalar heavy-heavy current
$\J_s=\Q_{s1}^+\Q_{s2}$ has the $\ga$-matrix structure (Fig.~\ref{FigCorr3})
\begin{equation}
K_s=A+B_1\v_1+B_2\v_2+C\v_2\v_1.
\label{Scal3}
\end{equation}
For the currents $\j_{P_2}$, $\overline{\j}_{P_1}$ with the definite parities
we have
\begin{equation}
K_{P_2 P_1}=A+P_1 B_1+P_2 B_2+P_1 P_2 C
=\frac{\frac14\Tr(1+P_1\v_1)(1+P_2\v_2)K_s}{1+P_1 P_2 v_1\cdot v_2}.
\label{ScalP3}
\end{equation}
Due to the linear relations~(\ref{Corr3}--\ref{Laplace3}),
the same $\ga$-matrix structures and relations between $K_s$ and $K_{P_2 P_1}$
hold in both the coordinate space and the momentum one, and also for
spectral densities. When calculating the correlator using the OPE,
even-dimensional terms contribute to $B_1$, $B_2$, and odd-dimensional ones%
---to $A$, $C$.

It is convenient to use the ``brick wall'' frame in which $\vec v_1=-\vec v_2$
is directed along $z$ for counting form factors. Angular momentum projection
onto $z$ is conserved: $j_{2z}=j_{1z}$. The reflection in any plane containing
$z$ transforms the state $|j,j_z{>}$ to $Pi^{2j}|j,-j_z{>}$. Therefore the
amplitude for $-j_{1z}$, $-j_{2z}$ is equal to that for $j_{1z}$, $j_{2z}$
up to a phase factor; the $0\to0$ transition is allowed only if the
``naturalness'' $P(-1)^j$ is conserved~\cite{Politzer}. For example,
$\frac12^+\to\frac12^+$, $\frac12^+\to\frac12^-$, and $\frac12^+\to\frac32^-$
transitions are described by one form factor each:
\begin{eqnarray}
\va{M,\frac12^+|\J_s|M,\frac12^+}&=&\xi(\ch\phi)\u_2 u_1,
\nonumber\\
\va{M,\frac12^-|\J_s|M,\frac12^+}&=&\tau_{1/2}(\ch\phi)\u_2\ga_5 u_1,
\label{ScalIW}\\
\va{M,\frac32^-|\J_s|M,\frac12^+}&=&\tau_{3/2}(\ch\phi)v_{1\mu}\u_{2\mu} u_1,
\nonumber
\end{eqnarray}
where $\ch\phi=v_1\cdot v_2$ is the cosine of the Minkovskian angle between
the world lines of the incoming and the outgoing heavy quark, and $u_\mu$ is
the Rarita-Schwinger wave function of the spin $\frac32$ meson. Here we have
slightly changed the notations as compared to~\cite{IW2}: in the original
notations right-hand sides of~(\ref{ScalIW}) should contain $2\tau_{1/2}$
and $\sqrt{3}\tau_{3/2}$. The contribution of $\frac12^+$ mesons to the
spectral density $\rho_s(\om_2,\om_1)$ is $\f_2 \f_1 \xi(\ch\phi)
\frac{1+\v_2}{2} \frac{1+\v_1}{2} \de(\om_2-\e_2) \om_1-\e_1)$, i.~e. the
contribution to $\rho_{++}(\om_2,\om_1)$ is
$\f_2 \f_1 \xi(\ch\phi) \de(\om_2-\e_2) \om_1-\e_1)$. The spectral density
of the correlator of the currents $\overline{\j_{s1,2}}u_{1,2}$ with some
specific polarizations $u_{1,2}$ is $\u_2 \rho_s(\om_2,\om_1) u_1$. The
contribution of $\frac12^+$ mesons to it is
$\f_2 \f_1 \u_2 \xi(\ch\phi) u_1 \de(\om_2-\e_2) \de(\om_1-\e_1)$.
Similar formulae hold for $\frac12^-$ mesons.

Now let's switch the heavy quark spin on. The correlator of
$\j_2=\overline{\Q}_2 \Ga_2 q$, $\overline{\j}_1=\q \G_1 \Q_1$,
and $\J=\overline{\Q}_1 \Ga \Q_2$ is (Fig.~\ref{FigCorr3})
\begin{equation}
K=\Tr\Ga_2\frac{1-\v_2}{2}\Ga\frac{1-\v_1}{2}\G_1K_s.
\label{Real3}
\end{equation}
In $K_s$, $\v_{1,2}$ may be replaced by $P_{1,2}=\pm1$ for $\Ga_{1,2}$ (anti-)
commuting with $\v_{1,2}$, and $K_s$ becomes the scalar function $K_{P_2P_1}$:
\begin{equation}
K=K_{P_2P_1}\Tr\Ga_2\frac{1-\v_2}{2}\Ga\frac{1-\v_1}{2}\G_1.
\label{RealP3}
\end{equation}
Let's introduce the currents $\q\M\Q$, $\M=\G\psi$ where $\psi$ is the spin
wave function, i.~e.\ $\M_{0^-}=\overline{\ga}_5=-\ga_5$,
$\M_{1^-}=\overline{\ga}_\mu e_\mu=\widehat{e}$. The spectral density of
their correlator is $\rho_{P_2P_1}\Tr\overline{\M}_2\frac{1-\v_2}{2}\Ga\M_1$;
the mesons' contribution to it is
$\f_{M_2}\f_{M_1}\va{M_2|\J|M_1}\de(\om_2-\e_2)\de(\om_1-\e_1)$. Hence we
obtain~\cite{IW,FGGW}
\begin{equation}
\va{M_2|\J|M_1}=\frac12\xi(\ch\phi)\Tr\overline{\M}_2\frac{1-\v_2}{2}\Ga
\frac{1-\v_1}{2}\M_1.
\label{RealIW}
\end{equation}
This formula expresses all form factors of transitions of a ground-state
$0^-$, $1^-$ meson to a ground-state $0^-$, $1^-$ meson under the action
of any heavy-heavy current $\J=\overline{\Q}_2\Ga\Q_1$ via one universal
Isgur-Wise form factor $\xi(\ch\phi)$. A similar formula with
$\tau_{1/2}(\ch\phi)$  holds for $\frac12^+\to\frac12^-$ transitions;
$\M=1$, $\widehat{e}\ga_5$ for $0^+$, $1^+$ mesons.

A three-point correlator at $\phi=0$ is expressed via the corresponding
two-point one:
\begin{eqnarray}
K(t_2,t_1)&=&\Pi(t_1+t_2),
\nonumber\\
K(\om_2,\om_1)&=&\frac{\Pi(\om_1)-\Pi(\om_2)}{\om_1-\om_2},
\label{Phi0}\\
\rho(\om_2,\om_1)&=&\rho(\om_1)\de(\om_2-\om_1).
\nonumber
\end{eqnarray}
The first two forms follow from each other by the Fourier transform.
The third form is necessary and sufficient for the first one because
of~(\ref{Disp2}); it can be obtained by the double backward Laplace
transform~(\ref{Laplace3}) of the single forward Laplace
transform~(\ref{Laplace2}) of the first form at imaginary times.
The third form can be also obtained from the second one by taking the
double discontinuity: the discontinuity of the first term in $\om_2$ is
$\pi i\de(\om_2-\om_1)\Pi(\om_1)$, and the discontinuity of this expression
in $\om_1$ is $\frac12(2\pi i)^2\rho(\om_1)\de(\om_2-\om_1)$; the second
term contributes equally.

To prove the first form, let's consider any diagram for the two-point
correlator in the coordinate space (for simplicity, with the scalar heavy
quark). Vertices along the heavy quark line have the times
$t_0<t_1<\cdots<t_{n-1}<t_n$, and the integration in $t_2,\ldots,t_{n-1}$
is performed. The integrand is an integral over coordinates of all vertices
not belonging to the heavy quark line. Now consider all diagrams for the
three-point correlator obtained by inserting the heavy-heavy vertex with
time $t$ (and $\phi=0$) to all the possible places along the heavy quark
line. These diagrams have the same integrand and the integration regions
$t_0<t_1<\cdots<t_{m-1}<t<t_m<\cdots<t_{n-1}<t_n$ ($m=1,\ldots,n$). These
regions span the whole integration region of the original diagram. Therefore
the sum of this set of three-point diagrams is equal to the two-point diagram.

\begin{figure}[p]
\begin{center}
\includegraphics[width=\linewidth]{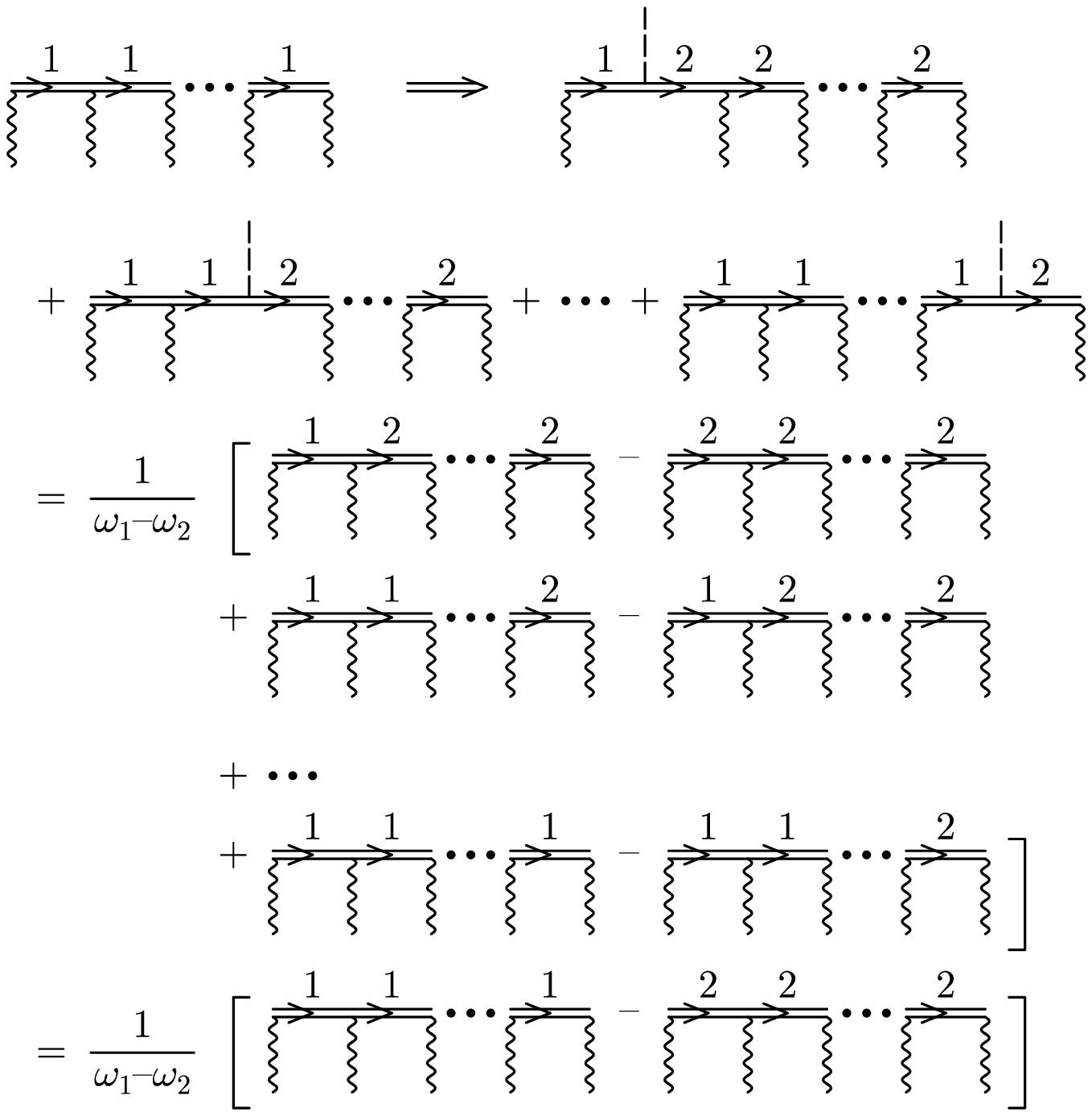}
\end{center}
\caption{Proof of the Ward identity. A digit 1 (2) near a heavy quark line
means that its energy includes $\om_1$ ($\om_2$)}
\label{FigWard}
\end{figure}

The second form can be easily proved in the momentum space in the exact
analogy with the QED Ward identity using the relation
$i\S(\om_1+\om')ii\S(\om_2+\om')=\frac{i\S(\om_1)-i\S(\om_2)}{\om_1-\om_2}$
(Fig.~\ref{FigWard}). In particular, it implies
$K(\om,\om)=\frac{d\Pi(\om)}{d\om}$.

Comparing the mesons' contributions to $\rho(\om_2,\om_1)$ and $\rho(\om)$,
we see
\begin{equation}
\xi(1)=1
\label{NormIW}
\end{equation}
for any $\frac12^+$ meson. For non-diagonal $\frac12^+\to\frac12^+$
transitions $\xi(1)=0$ because $\rho(\om_2,\om_1)=0$ off the diagonal.
The physical meaning of this is simple: when the current $\J$ replaces
the old heavy quark by the new one with the same velocity and color,
light fields don't notice it. The formulae for the form factors at $\phi=0$
equivalent to~(\ref{RealIW}, \ref{NormIW}) were first proposed in the
quark model framework~\cite{QuarkModel}.

The variable $\ch\phi$ is related to the momentum transfer $q^2$ (for
simplicity in the case of equal heavy quark masses) by the formula
$q^2=2m^2(1-\ch\phi)$. Form factors are analytic functions of $q^2$ with
the cut in the annihilation channel from $4m^2$ to $+\infty$. Therefore
the Isgur-Wise function $\xi(\ch\phi)$ is an analytic function in the
$\ch\phi$ plane with the cut from $-1$ to $-\infty$. Geometrically speaking,
$ch\phi>1$ corresponds to Minkovskian angles between the incoming and
outgoing heavy quark world lines (scattering or decay); $\ch\phi=1$ means
the straight world line---no transition at all; nothing special happens at
$\ch\phi<1$, only the angle becomes Euclidean; $\ch\phi=-1$ is really
a singular point where the quark returns along the same world line;
$\ch\phi<-1$ corresponds again to Minkovskian angles only one of the world
lines is directed to the past (annihilation).

In the rest of this Section, we shall for simplicity live in the world
with the scalar $b$ quark decaying into the scalar $c$ quark plus the scalar
$W$ boson. Their masses can be adjusted in such a way as to give any desired
$\ch\phi$. The quark decay matrix element is simply $M=g$ where $g$ is the
coupling constant.

Until now we discussed exclusive decays of the $\frac12^+$ $B$ meson.
Inclusive decays can be also treated in HQET~\cite{Bjorken,Bjorken2}.
The matrix element of the decay $B\to X+W$ (where $X$ is any hadronic state
containing the $c$ quark) has the structure $M=g\overline{\psi}_2(X)u_1$.
Its spin-averaged square is $\overline{|M|^2}=\frac{g^2}{2}\Tr\frac{1+\v_1}{2}
\psi_2(X)\overline{\psi}_2(X)$. Let's sum over hadronic states $X$ with
the energy $\e$:
\begin{eqnarray}
&&W(\e,\ch\phi)=\sum_X\psi_2(X)\overline{\psi}_2(X)\de(\e_X-\e)
=a+b\v_2,
\nonumber\\
&&d\Ga=d\Ga_0 w(\e,\ch\phi)d\e,
\label{StructFun}\\
&&w(\e,\ch\phi)=\frac12\Tr\frac{1+\v_1}{2}W(\e,\ch\phi)=a+b\ch\phi.
\nonumber
\end{eqnarray}
The meson decay rate $d\Ga$ differs from the quark decay rate $d\Ga_0$ by
the structure function $w(\e,\ch\phi)$ where $\e$ is the energy of the
hadronic state $X$ (minus the $c$ quark mass) in the $v_2$ rest frame.
The quark-hadron duality tells
us that the total meson decay rate is equal to the quark one. This is the
Bjorken sum rule~\cite{Bjorken,Bjorken2}
\begin{equation}
\int\limits_0^\infty w(\e,\ch\phi)d\e=1.
\label{Bjorken}
\end{equation}
The second sum rule follows from the momentum conservation~\cite{Burkardt}.
The initial ground-state meson has the energy $\e_g$ in the $v_1$ rest frame.
In the $v_2$ rest frame this corresponds to the energy $\e_g\ch\phi$ plus
an irrelevant momentum orthogonal to $v_2$. Therefore the average energy
of the hadronic system $X$ must be
\begin{equation}
\int\limits_0^\infty w(\e,\ch\phi)\e d\e=\e_g\ch\phi.
\label{Burkardt}
\end{equation}
Inclusive semileptonic $B$ decays in HQET were also discussed in~\cite{Chay}.

Now we shall explicitly write down some contribution to this sum rule.
The spin-averaged matrix elements squared for the decays
$\frac12^+\to\frac12^+$, $\frac12^+\to\frac12^-$, and
$\frac12^+\to\frac32^-$~(\ref{ScalIW}) are
\begin{eqnarray}
\overline{|M|^2_{\frac12^+}}&=&\frac{g^2\xi^2}{2}
\Tr\frac{1+\v_1}{2}\frac{1+\v_2}{2}=g^2\xi^2\frac{\ch\phi+1}{2},
\nonumber\\
\overline{|M|^2_{\frac12^-}}&=&\frac{g^2\tau_{1/2}^2}{2}
\Tr\frac{1+\v_1}{2}\ga_5\frac{1+\v_2}{2}\overline{\ga}_5
=g^2\tau_{1/2}^2\frac{\ch\phi-1}{2},
\label{ScalM2}\\
\overline{|M|^2_{\frac32^-}}&=&g^2\tau_{3/2}^2
\frac{(\ch\phi+1)^2(\ch\phi-1)}{3}.
\nonumber
\end{eqnarray}
Here we have used the Rarita-Schwinger density matrix.
The decay $\frac12^+\to\frac12^+$ is $S$-wave, hence $\overline{|M|^2}$
is constant at $\phi\to0$. The decays $\frac12^+\to\frac12^-$,
$\frac12^+\to\frac32^-$ are $P$-wave, hence $\overline{|M|^2}\sim\phi^2$.
The decays to the $D$-wave mesons $\frac32^+$, $\frac52^+$ are $D$-wave
with $\overline{|M|^2}\sim\phi^4$, etc. Matrix elements squared in the
annihilation channel $W\to\overline{B}D$ differ from~(\ref{ScalM2}) by
the absence of the factor $\frac12$ coming from the averaging over the
initial meson spin states. The first decay $W\to\frac12^-\frac12^+$ is
$P$-wave, the second one $W\to\frac12^-\frac12^-$ is $S$-wave, and the third
one $W\to\frac12^-\frac32^-$ is $D$-wave. This determines the threshold
behavior of~(\ref{ScalM2}) at $\ch\phi\to-1$.

Therefore the Bjorken sum rule reads~\cite{IW2}
\begin{equation}
\sum_{\frac12^+}\xi^2\frac{\ch\phi+1}{2}
+\sum_{\frac12^-}\tau_{1/2}^2\frac{\ch\phi-1}{2}
+\sum_{\frac32^-}\tau_{3/2}^2\frac{(\ch\phi+1)^2(\ch\phi-1)}{3}
+\cdots=1,
\label{BjorkenIW}
\end{equation}
where the sums are over resonances in the $j^P$ channels, and dots mean
contributions of other channels (see~\cite{HighMeson}). The Burkardt sum
rule~(\ref{Burkardt}) has the similar form. The simplest consequence
of~(\ref{BjorkenIW}) is that the decay rate to the ground-state $\frac12^+$
meson is less than the total one:
\begin{equation}
\xi(\ch\phi)\le\sqrt{\frac{2}{\ch\phi+1}}.
\label{BoundIW}
\end{equation}
Of course, at $\ch\phi\gg1$ most decays are inelastic, and
$\xi(\ch\phi)\ll1/\sqrt{\ch\phi}$.

Let's consider the Bjorken sum rule~(\ref{BjorkenIW}) at small $\ch\phi-1$,
and expand it to the linear terms. The channels denoted by dots don't
contribute because they are at least $D$-wave. Higher resonances in the
$\frac12^+$ channel don't contribute because they have
$\xi(\ch\phi)=O(\ch\phi-1)$. We are left with~\cite{IW2}
\begin{equation}
\xi'(1)=-\frac14-\frac14\sum_{\frac12^-}\tau_{1/2}^2(1)
-\frac23\sum_{\frac32^-}\tau_{3/2}^2(1).
\label{SlopeIW}
\end{equation}
This gives us the Bjorken bound $\xi'(1)<-\frac14$ (evident also from%
~(\ref{BoundIW})). Similarly, the Burkardt sum rule~(\ref{Burkardt}) leads
to the optical (Thomas-Reiche-Kuhn) sum rule~\cite{Voloshin}
\begin{equation}
\frac14\sum_{\frac12^-}(\e_{1/2}-\e_g)\tau_{1/2}^2(1)
+\frac23\sum_{\frac32^-}(\e_{3/2}-\e_g)\tau_{3/2}^2(1)=\frac12\e_g.
\label{Optical}
\end{equation}
It can be used for obtaining bounds on $\xi'(1)$~\cite{Voloshin}.

It is also possible to establish the bound on the Isgur-Wise form factor
at the cut~\cite{dRT}: the decay rate $W\to\overline{B}D$ is less than
the total decay rate $W\to\overline{b}c$. The meson decay rate for each
flavour is given by~(\ref{ScalM2}) without the spin-averaging factor
$\frac12$; the quark decay rate is $\overline{|M|^2}=g^2N_c$ where $N_c$
is the number of colors. If there are $n_l$ light flavours for which
$\xi(\ch\phi)$ is approximately the same, then
\begin{equation}
n_l |\xi(\ch\phi)|^2|\ch\phi+1|\le N_c.
\label{CutBound}
\end{equation}
In general the left-hand side is the sum over light flavours.
The factor $n_l$ was erroneously omitted in~\cite{dRT}; the phrase justifying
this seems to have no sense. At $|\ch\phi|\gg1$, the $B\overline{D}$ channel
constitutes a small fraction of the total $W\to b\overline{c}$ width, and
$|\xi(\ch\phi)|\ll1/\sqrt{|\ch\phi|}$. One could include also higher states'
contributions in the left-hand side; this should be done with caution because
of the possibility of double counting. The inequality~(\ref{CutBound}) is
applicable only sufficiently far from the threshold, at
$|\ch\phi+1|\gg\pi^2\al_s^2$. Near the threshold the Coulomb interaction
between the heavy quark and antiquark is essential. The total decay width
on the right-hand side is not equal to its free-quark value $N_c$;
it contains high narrow resonances at the quarkonium levels. Moreover,
the very concept of the Isgur-Wise form factor is inapplicable in this region.
The HQET picture is based on the fact that heavy quarks move along straight
world lines, but at velocities $\sim\pi\al_s$ they really rotate around
each other.

If the inequality~(\ref{CutBound}) were true everywhere on the cut, we would
immediately arrive at a paradox~\cite{Broadhurst}. Consider the function
$f(\ch\phi)$ analytic in the $\ch\phi$ plane with the cut from $-1$ to
$-\infty$. On the cut $|f|^2\le\frac{N_c}{2n_l}$, and $f(1)=1$. This is
consistent with the maximum modulus theorem only at
\begin{equation}
2n_l\le N_c
\label{paradox}
\end{equation}
what is not the case in our world. The more detailed analysis%
~\cite{Broadhurst} shows that it is possible to obtain similar inequalities
(with the constant smaller than 2) using weight functions that are rather
insensitive to the threshold region, and the paradox remains.

HQET can also be used for description of heavy to light transitions%
~\cite{HeavyLight} and rare $B$ decays~\cite{Rare}. In these cases
the heavy quark spin symmetry is not so restrictive, and more form factors
are necessary. Relations between $B$ and $D$ decays can be established using
the heavy quark spin-flavour symmetry and the isospin symmetry. Inclusive
heavy to light decays are considered in~\cite{Bjorken2}; they are described
by several structure functions obeying sum rules in the deep inelastic
region.

An interesting approach in which the parameter $\left(\frac{\phi_{\rm max}}{2}
\right)^2=\left(\frac{m_b-m_c}{m_b+m_c}\right)^2$
is considered small was proposed in~\cite{ShifVol2}. To the leading order in
this parameter, the quark-hadron duality is perfect: the $b\to cW$ quark
decay rate is equal to the $B\to DW$ meson decay rate (see~(\ref{BjorkenIW})).
This is true for all heavy quark polarizations; in particular, the hadronic
tensor $\va{B|\j|X}\va{X|\j^+|B}$ summed over ground state mesons $X=D$,
$D^*$ is equal to the corresponding quark tensor summed over $c$ polarizations.

\section{Baryons}
\label{SecBaryon} \setcounter{equation}{0}

For ground-state baryons, the light quark spins can add giving $j^P=0^+$ or
$1^+$. In the first case their spin wave function is antisymmetric, the Fermi
statistics and the antisymmetry in color require an antisymmetric flavour
wave function. Hence the light quarks must be different; if they are $u$, $d$
then their isospin $I=0$. With the heavy quark spin switched off, we have the
$0^+$ $I=0$ baryon $\La_Q$. If one of the light quarks is $s$, we obtain the
isodoublet $\Xi'_Q$ forming together with $\La_Q$ the $SU(3)$ antitriplet.
In the $1^+$ case the flavour wave function is symmetric; if the light quarks
are $u$, $d$ then their isospin $I=1$. So we have the $1^+$ isotriplet $\Si_Q$;
with one $s$ quark---the isodoublet $\Xi_Q$; with two $s$ quarks---the
isosinglet $\Om_Q$. Together they form the $SU(3)$ sextet. With the heavy
quark spin switched on, the scalar baryons $\La_Q$, $\Xi'_Q$ become
$\frac12^+$; the vector baryons form degenerate $\frac12^+$, $\frac32^+$
doublets $\Si_Q$, $\Si^*_Q$; $\Xi_Q$, $\Xi^*_Q$; $\Om_Q$, $\Om^*_Q$.

Baryons and their form factors in HQET were considered
in~\cite{IWb,MRR,Mainz}. Two-point and three-point HQET sum rules
were investigated in~\cite{GrYak}.

\begin{sloppypar}
Baryon currents with the scalar heavy quark have the form
$\j_s=\ep^{abc}(q^{Ta}C\Ga\tau q^b)\Q^c_s$ where $q^T$ means $q$ transposed
and $C$ is the charge conjugation matrix (because $q^T C$ is transformed like
$\q$ under the action of the Lorentz group). Here $\tau$ is a flavour matrix,
symmetric for $0^+$ baryons and antisymmetric for $1^+$ ones. We shall
abbreviate it to $\j_s=(q^T C\Ga q)\Q_s$. A light quark pair with $j^P=0^+$
corresponds to the current $a=q^T C\ga_5 q$, and with $1^+$---to
$\vec a=q^T C\vec\ga q$ (one can easily check it using the $P$-conjugation
$q\to\ga_0 q$). It is also possible to insert $\ga_0$ into these currents
without changing their quantum numbers. So, the scalar heavy quark currents
with the quantum numbers of $\La_Q$, $\Si_Q$ are $\j_{\La s}=a\Q_s$,
$\vec{\j}_{\Si s}=\vec{a}\Q_s$.
\end{sloppypar}

With the real-world spin $\frac12$ heavy quark, the current $\j=a\Q$ has the
spin $\frac12$; the current $\vec{\j}=\vec{a}\Q$ contains spin $\frac12$ and
spin $\frac32$ components. The part
$\vec{\j}_{3/2}=\vec{\j}+\frac13\vec{\ga}\,\vec{\ga}\cdot\vec{\j}$ satisfies
the condition $\vec{\ga}\cdot\vec{\j}_{3/2}=0$ and hence has the spin
$\frac32$. The other part $\vec{\j}_{1/2}=-\frac13\vec{\ga}\,
\vec{\ga}\cdot\vec{\j}=\frac13\vec{\ga}\ga_5 \j_{1/2}$,
$\j_{1/2}=\vec{a}\cdot\vec{\ga}\,\ga_5\Q$ has the spin $\frac12$. Finally we
obtain the currents $\j=(q^T C\Ga q)\Ga'\Q$ with the quantum numbers of
$\La_Q$, $\Si_Q$, $\Si^*_Q$~\cite{GrYak}
\begin{equation}
\j_\La=(q^T C\ga_5 q)\Q,
\quad
\j_\Si=(q^T C\vec\ga q)\cdot\vec\ga\,\ga_5\Q,
\quad
\vec{\j}_{\Si^*}
=(q^T C\vec\ga q)\Q+\frac13\vec\ga(q^T C\vec\ga q)\cdot\vec\ga\Q,
\label{BarCur}
\end{equation}
and similar currents with the extra $\ga_0$ inside the brackets.

\begin{figure}[ht]
\begin{center}
\includegraphics{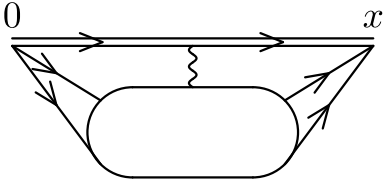}
\end{center}
\caption{Correlator of two HQET baryonic currents}
\label{FigBcorr2}
\end{figure}

The correlators of two baryon currents with the scalar heavy quark have the
structure (Fig.~\ref{FigBcorr2})
\begin{eqnarray}
i\va{T\j_{\La s}(x)\j^+_{\La s}(0)}&=&\de(\vec x) \Tr\tau^+\tau \Pi_\La(x_0),
\label{Bscal2}\\
i\va{T\j_{\Si si}(x)\j^+_{\Si sj}(0)}&=&\de_{ij}\de(\vec x) \Tr\tau^+\tau
\Pi_\Si(x_0).
\nonumber
\end{eqnarray}
From now on we shall for simplicity assume the normalization $\Tr\tau^+\tau=1$.
If we denote $\va{0|\j_\La|\La_Q,0^+}=\f_{\La,0^+}$,
$\va{0|\vec{\j}_\Si|\Si_Q,1^+}=\f_{\Si,1^+}\vec{e}$, then the baryon
contribution to $\rho_{\La,\Si}(\om)$ is
$\f^2_{\La,\Si}\de(\om-\e_{\La,\Si})$.

\begin{sloppypar}
Now let's switch the heavy quark spin on. The correlators are
(Fig.~\ref{FigBcorr2})
\begin{equation}
\Pi=\left(\Ga'_1\frac{1+\ga_0}{2}\G'_2\right)\Pi_s,
\label{Breal2}
\end{equation}
where tensor indices may be contracted between $\Pi_s$ and $\Ga'_{1,2}$.
The same relation holds for $\Pi(t)$, $\Pi(\om)$, and $\rho(\om)$. For
$\La_Q$, $\Si_Q$, $\Si^*_Q$ we obtain
\begin{eqnarray}
\rho_\La&=&\frac{1+\ga_0}{2}\rho_{\La s},
\nonumber\\
\rho_\Si&=&\ga_i\ga_5\frac{1+\ga_0}{2}\ga_j\ga_5\de_{ij}\rho_{\Si s}
=3\frac{1+\ga_0}{2}\rho_{\Si s},
\label{BrealS2}\\
\rho_{\Si^*}&=&\left(\de_{ii'}-\frac13\ga_i\ga_{i'}\right)\frac{1+\ga_0}{2}
\left(\de_{jj'}-\frac13\ga_j\ga_{j'}\right)\de_{i'j'}\rho_{\Si s}
\nonumber\\
&=&\frac{1+\ga_0}{2}\left(\de_{ij}+\frac13\ga_i\ga_j\right)\rho_{\Si s}.
\nonumber
\end{eqnarray}
If we denote $\va{0|\j_\La|\La_Q,\frac12^+}=\f_{\La,\frac12^+}u$,
$\va{0|\j_\Si|\Si_Q,\frac12^+}=\f_{\Si,\frac12^+}u$,
$\va{0|\vec{\j}_{\Si^*}|\Si^*_Q,\frac32^+}=\f_{\Si^*,\frac32^+}\vec{u}$,
then the baryon contributions to~(\ref{BrealS2}) are
$\frac{1+\ga_0}{2}\f_{\La,\frac12^+}^2\de(\om-\e_\La)$,
$\frac{1+\ga_0}{2}\f_{\Si,\frac12^+}^2\de(\om-\e_\Si)$,
$\frac{1+\ga_0}{2}\left(\de_{ij}+\frac13\ga_i\ga_j\right)
\f_{\Si^*,\frac32^+}^2\de(\om-\e_\Si)$. $\La_Q$ is a spin symmetry singlet,
therefore there are no interesting predictions of the spin symmetry in this
channel. Baryons in $\Si_Q$ and $\Si^*_Q$ channels are degenerate:
$\e_{\Si,\frac12^+}=\e_{\Si^*,\frac32^+}=\e_{\Si,1^+}$, and
$\frac1{\sqrt{3}}\f_{\Si,\frac12^+}=\f_{\Si^*,\frac32^+}=\f_{\Si,1^+}$.
Note that both sides of the definitions $\va{0|j|B}=f_B u$ get the same factor
$\sqrt{2m}$ when going from the relativistic normalization to the
nonrelativistic one. Therefore the QCD quantities $f_B=\f_B$ don't depend on
$m$ (compare with~(\ref{Scaling})).
\end{sloppypar}

\begin{figure}[ht]
\begin{center}
\includegraphics{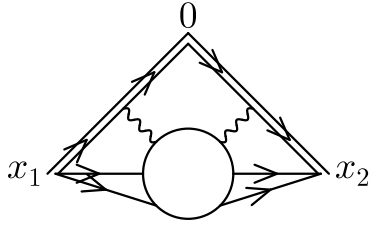}
\end{center}
\caption{Correlator of two HQET baryonic currents and a heavy-heavy current}
\label{FigBcorr3}
\end{figure}

\begin{sloppypar}
The correlators of two baryon currents with the scalar heavy quark and the
scalar heavy-heavy current $\J_s$ have the structure (Fig.~\ref{FigBcorr3})
\begin{eqnarray}
&&i^2\va{T\j_{\La s}(x_2)\J_s(0)\j^+_{\La s}(x_2)}=\Tr\tau^+\tau
\nonumber\\&&\quad
\int\limits_0^\infty dt_2\de(x_2-v_2t_2)
\int\limits_0^\infty dt_1\de(x_1+v_1t12)
K_\La(t_2,t_1),
\label{Bscal3}\\
&&i^2\va{T\j_{\Si s\mu}(x_2)\J_s(0)\j^+_{\Si s\nu}(x_2)}=\Tr\tau^+\tau
\nonumber\\&&\quad
\int\limits_0^\infty dt_2\de(x_2-v_2t_2)
\int\limits_0^\infty dt_1\de(x_1+v_1t12)
K_{\Si\mu\nu}(t_2,t_1),
\nonumber\\&&\quad
K_{\Si\mu\nu}=K_{\Si||}e_{2||\mu}e_{1||\nu}+K_{\Si\bot}\de_{\bot\mu\nu},
\nonumber
\end{eqnarray}
where $e_{1||}=(v_2-\ch\phi v_1)/\sh\phi$, $e_{2||}=-(v_1-\ch\phi v_2)/\sh\phi$
are the $\Si_Q$ polarization vectors in the scattering plane,
$\de_{\bot\mu\nu}=\sum e_{\bot\mu}e_{\bot\nu}=\left[\ch\phi(v_{1\mu}v_{2\nu}
+v_{2\mu}v_{1\nu})-v_{1\mu}v_{1\nu}-v_{2\mu}v_{2\nu}\right]/\sh^2\phi
-g_{\mu\nu}$.
\end{sloppypar}

\begin{sloppypar}
According to the rules~\cite{Politzer}, the transition $\La_Q\to\La_Q$ is
described by one form factor ($j_{1z}=j_{2z}=0$); $\La_Q\to\Si_Q$ is forbidden
by naturalness; $\Si_Q\to\Si_Q$ is described by two form factors
($j_{1z}=j_{2z}=0$ and $\pm1$):
\begin{eqnarray}
\va{\La_Q|\J_s|\La_Q}&=&\xi_\La(\ch\phi),
\label{BscalIW}\\
\va{\Si_Q|\J_s|\Si_Q}&=&\xi_{\Si\mu\nu}e^*_{2\mu}e_{1\nu},
\quad
\xi_{\Si\mu\nu}=\xi_{\Si||}(\ch\phi)e_{2||\mu}e_{1||\nu}
+\xi_{\Si\bot}(\ch\phi)\de_{\bot\mu\nu}.
\nonumber
\end{eqnarray}
The contribution of $\La_Q$, $\Si_Q$ to $\rho_{\La,\Si||,\Si\bot}(\om_2,\om_1)$
is $\f_{\La,\Si}^2\xi_{\La,\Si||,\Si\bot}\de(\om_2-\e_{\La,\Si})
\de(\om_1-\e_{\La,\Si})$. The spectral density of the correlator of the
currents $\j^+_{\Si\mu}e_\mu$ with some specific polarizations $e_{1,2}$
is $\rho_{\Si\mu\nu}e^*_{2\mu}e_{1\nu}$; the $\Si_Q$ contribution to it is
$\f_\Si^2 \xi_{\Si\mu\nu}e^*_{2\mu}e_{1\nu}$.
\end{sloppypar}

Now let's switch the heavy quark spin on. The correlators are
(Fig.~\ref{FigBcorr3})
\begin{equation}
K=\left(\Ga'_2\frac{1+\v_2}{2}\Ga\frac{1+\v_1}{2}\G'_1\right)K_s
\label{Breal3}
\end{equation}
Let's introduce the currents $\overline{\j}u=(\q\G C^{-1}\q^T)\overline{\Q}\B$,
$\B=\G'u$. Rewriting~(\ref{BarCur}) in the covariant form
$\j_\Si=(Q^T C\ga_\mu q)\Ga'_\mu \Q$,
$\j_{\Si^*\nu}=(Q^T C\ga_\mu q)\Ga'_{\mu\nu}\Q$, we obtain $\B_\La=u$,
$\B_{\Si\mu}=-(\ga_\mu+v_\mu)u$, $\B_{\Si^*\mu}=u_\mu$.
The spectral density of the correlator is
$\left(\overline{\B}_2\frac{1+\v_2}{2}\Ga\frac{1+\v_2}{2}\B_1\right)\rho_s$
(where tensor indices may be contracted between $\rho_s$ and $\B_{1,2}$);
the baryons' contribution to it is
$\f_{B_2}\f_{B_1}\va{B_2|\J|B_1}\de(\om_2-\e_2)\de(\om_1-\e_1)$. Hence we
obtain~\cite{IWb,MRR,Mainz}
\begin{eqnarray}
\va{\La_Q|\J|\La_Q}&=&\xi_\La(\ch\phi)\overline{u}_2\Ga u_1,
\nonumber\\
\va{\Si_Q|\J|\Si_Q}&=&\frac13\xi_{\Si\mu\nu}
\overline{u}_2(\ga_\mu+v_{2\mu})\ga_5\Ga(\ga_\nu-v_{1\nu})\ga_5 u_1,
\nonumber\\
\va{\Si^*_Q|\J|\Si_Q}&=&\frac1{\sqrt3}\xi_{\Si\mu\nu}
\overline{u}_{2\mu}\Ga(\ga_\nu-v_{1\nu})\ga_5 u_1,
\label{BrealIW}\\
\va{\Si^*_Q|\J|\Si^*_Q}&=&\xi_{\Si\mu\nu}
\overline{u}_{2\mu}\Ga u_{1\nu}.
\nonumber
\end{eqnarray}
The result for $\La_Q$ is particularly simple because light fields have
$j^P=0^+$, and the spin of $\La_Q$ is carried by the heavy quark.

At the point $\phi=0$
\begin{equation}
\xi_\La(1)=\xi_{\Si||}(1)=\xi_{\Si\bot}(1)=1
\label{BnormIW}
\end{equation}
(at this point $\xi_{\Si ij}=\de_{ij}$ because there are no selected
directions).

Inclusive $\La_Q$ decays were treated in~\cite{Bjorken,Bjorken2}. With the
scalar heavy quarks, the matrix element of the decay $\La_Q\to X$ has the
structure $M=g\phi_2^*(X)\phi_1$ where $phi_1$ is the scalar $\La_Q$ wave
function. The $\La_Q$ decay rate is given by~(\ref{StructFun}) with the
structure function $w(\e,\ch\phi)=\sum_X\phi_2^*(X)\phi_2(X)\de(\e_X-\e)$
obeying the Bjorken sum rule~(\ref{Bjorken}). Transitions to the
excited baryons and their contribution to the Bjorken sum rule were considered
in~\cite{IWY,HighBaryon}.

Polarization effects in $\La_Q$ decays were discussed in~\cite{Polar}.
Heavy to light baryon transitions were considered in~\cite{MRR,Mainz}; they
are described by several form factors. Inclusive heavy to light decays and
sum rules for them in the deep inelastic region were investigated in%
~\cite{Bjorken2}. At a large number of colors, baryons are bound states
of a chiral soliton and a meson; this model was considered in~\cite{Soliton}.

\section{Renormalization}
\label{SecRenorm} \setcounter{equation}{0}

Renormalization properties (anomalous dimensions etc.) of HQET are different
from that of QCD. The ultraviolet behavior of a HQET diagram is determined
by the region of loop momenta much larger than all characteristic scales
of the process but much less than the heavy quark mass which tends to infinity
from the very beginning. It has nothing to do with the ultraviolet behavior
of the corresponding QCD diagram with the heavy quark line which is
determined by the region of loop momenta much larger than the heavy quark
mass. In the conventional QCD the first region produces hybrid logarithms%
~\cite{ShifVol1,PolitzerWise}, and the problem of summation of these
logarithmic corrections is highly nontrivial. In HQET hybrid logarithms
become ultraviolet logarithmic divergences governed by the renormalization
group with corresponding anomalous dimensions. For example, correlators of
QCD meson currents contain large hybrid logarithmic corrections
$\al_s\log\frac{m}{\om}$. Correlators of the corresponding HQET currents
contain instead corrections $\al_s\log\frac{\om}{\mu}$ where $\mu$ is
the normalization point. The dependence on $\mu$ is determined by the
currents' anomalous dimensions; the corrections are small at $\mu\sim\om$.

HQET is closely related to the theory of Wilson lines in QCD~\cite{Polyakov}.
As follows from the Lagrangian~(\ref{HQETlagr}), the static
quark propagator in a gluon field is the straight Wilson line
\begin{equation}
\S(x) = -i\te(x_0)\de(\vec x) P\exp ig\int A_\mu dx_\mu.
\label{WilsonLine}
\end{equation}
The effective Lagrangian identical to the HQET
Lagrangian~(\ref{CovarLagr}) (strictly speaking, with the scalar
static quark Lagrangian~(\ref{SuperLagr})) was proposed
in~\cite{GervaisNeveu} for investigation of Wilson lines.
Their renormalization properties were considered in~\cite{WilsonLine}.

One-loop renormalization of straight Wilson lines (static quark propagators)
and cusps on them (heavy-heavy velocity changing currents) are known from%
~\cite{Polyakov}. Two-loop calculation~\cite{Aoyama} for straight Wilson
lines is incorrect; the correct result was obtained in~\cite{Knauss}.
It was also obtained in~\cite{BroadGraySchilcher} starting from the
on-shell renormalization of the QCD heavy quark propagator at finite $m$,
and in~\cite{BroadGr1,Gimenez} in the HQET framework. Two-loop renormalization
of a cusp on a Wilson line was first considered in~\cite{Knauss}, but the
authors were unable to get rid of all double integrals. A simple result
containing only simple single integrals was obtained in~\cite{Korchemsky}.
The attempt~\cite{Ji} in the HQET framework was unsuccessful: the result
contains a double integral (with a variable undefined in the paper); some
other integrals are in fact equal to 0 or each other.

One-loop renormalization of the heavy-light bilinear current in HQET was
first considered in~\cite{ShifVol1,PolitzerWise}. Two-loop corrections were
obtained in~\cite{BroadGr1,Gimenez} (in the second paper, a different
external momentum configuration was chosen which made calculations more
difficult). Four-quark operators with two static quark fields were also
investigated in~\cite{Gimenez}. One-loop renormalization of baryon currents
was considered in~\cite{GrYak}.

We shall use \MS{} scheme, the space dimension $D=4-2\ep$. The HQET
lagrangian~(\ref{HQETlagr}) expressed via bare fields and couplings is
\begin{eqnarray}
L&=&\Q_b^+i(\partial-ig_b A^a_b t^a)_0 \Q_b
+ \q_b i(\widehat{\partial}-ig_b\widehat{A}^a_b t^a) q
\nonumber\\
&-&\frac14 G^a_{b\mu\nu}G^a_{b\mu\nu}
+ \frac1{2a_b}(\partial_\mu A^a_{b\mu})^2 + ({\rm ghost}).
\label{BareLagr}
\end{eqnarray}
The bare quantities are related to the renormalized ones as
\begin{eqnarray}
&&\Q_b=\m^{-\ep}\Z_Q^{1/2}\Q,\quad
q_b=\m^{-\ep}Z_q^{1/2}q,\quad
A^a_{b\mu}=\m^{-\ep}A^a_\mu,
\nonumber\\
&&g_b=\m^\ep Z_\al^{1/2}g,\quad
a_b=Z_A a,
\label{BareRenorm}
\end{eqnarray}
where $Z_q$, $Z_A$, $Z_\al$ are the same as in QCD with $n_l$ light flavours
(there are no static quark loops), and $\m^2=\mu^2 e^\ga/4\pi$, $\mu$ is the
normalization point. The static quark field renormalization constant $\Z_Q$
is determined from the requirement that the renormalized propagator
$\S(\om)=\S_b(\om)/\Z_Q$ is finite. If we denote the sum of bare
one-particle-irreducible static quark diagrams $-i\widetilde{\Si}_b(\om)$,
then the propagator
$\S_b(\om)=\S_0(\om)+\S_0(\om)\widetilde{\Si}_b(\om)\S_0(\om)+\cdots
=1/(\om-\widetilde{\Si}_b(\om))$.

\begin{figure}[ht]
\begin{center}
\includegraphics{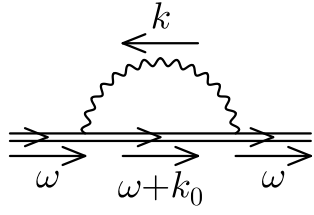}
\end{center}
\caption{HQET mass operator}
\label{FigMassOp}
\end{figure}

The bare one-loop HQET mass operator (Fig.~\ref{FigMassOp}) in the Feynman
gauge $a_b=0$ is
\begin{equation}
\widetilde{\Si}(\om)=-iC_F g_b^2\int\frac{d^D k}{(2\pi)^D}\frac1{k^2(\om+k_0)}
\label{MassOp}
\end{equation}
where $C_F=\frac{N_c^2-1}{2N_c}$.
A variant of the Feynman parametrization
\begin{equation}
\frac1{a^\al b^\be}=\frac{\Ga(\al+\be)}{\Ga(\al)\Ga(\be)}
\int\limits_0^\infty\frac{y^{\be-1}dy}{(a+yb)^{\al+\be}}
\label{Feynman}
\end{equation}
is used to combine a square denominator $a$ with a linear denominator $b$;
the parameter $y$ has the dimension of mass. We have
\begin{equation}
\widetilde{\Si}(\om)=-iC_F g_b^2\int\limits_0^\infty dy
\int\frac{d^D k}{(2\pi)^D}\frac1{(k^2+yk_0+y\om)^2}.
\label{MassOp2}
\end{equation}
The denominator is equal to $k^{\prime2}-\frac{y^2}{4}+\om y$,
$k'=k+\frac{y}{2}v$. Using the standard formula
\begin{equation}
\int\frac{d^D k}{(2\pi)^D}\frac1{(k^2-a^2)^n}=
\frac{i(-1)^n\Ga(n-\frac{D}{2})(a^2)^{D/2-n}}{(n-1)!(4\pi)^{D/2}},
\label{LoopIntegral}
\end{equation}
we obtain
\begin{equation}
\widetilde{\Si}(\om)=\frac{C_F g_b^2}{(4\pi)^{D/2}}
\Ga\left(2-{\textstyle\frac{D}{2}}\right)
\int\limits_0^\infty\left(\frac{y^2}{4}-\om y\right)^{D/2-2}dy.
\label{MassOp3}
\end{equation}
The integral
\begin{equation}
\int\limits_0^\infty y^\al(ay+b)^\be dy
=\frac{b^{\al+\be+1}}{a^{\al+1}}\frac{\Ga(-1-\al-\be)\Ga(1+\al)}{\Ga(-\be)}
\label{IntY}
\end{equation}
is calculated using the substitution $y=\frac{b}{a}\left(\frac1z-1\right)$.
Finally,
\begin{equation}
\widetilde{\Si}(\om)=\frac{C_F g_b^2}{(4\pi)^{D/2}}
2(-2\om)^{D-3}\Ga(3-D)\Ga\left({\textstyle\frac{D}{2}}-1\right).
\label{MassOpRes}
\end{equation}
Requiring the finiteness of $\S(\om)=\S_b(\om)/\Z_Q$ with a minimal
$\Z_Q=1+c\frac{\al_s}{\ep}$, we find
\begin{equation}
\Z_Q=1+C_F\frac{\al_s}{2\pi\ep}.
\label{ZQ}
\end{equation}

\begin{figure}[ht]
\begin{center}
\includegraphics{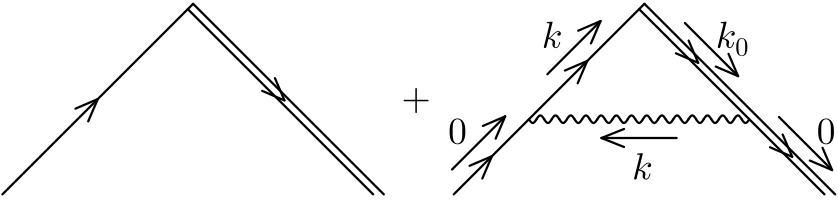}
\end{center}
\caption{Heavy-light vertex}
\label{FigHL}
\end{figure}

Now we shall consider renormalization of a heavy-light bilinear current.
The bare current $\j_b=\overline{\Q}_b\Ga q_b$ is related to the renormalized
one as $\j_b=\m^{-2\ep}\Z_j\j$, or $\overline{\Q}\Ga q=\Z_\Ga\j$ where
$\Z_j=Z_q^{1/2}\Z_Q^{1/2}\Z_\Ga$. Then the matrix element
$\widetilde{\Ga}=\va{\Q|\overline{\Q}\Ga q|q}=\Z_\Ga\va{\Q|\j|q}$ where
the matrix element of $\j$ is finite. The vertex $\widetilde{\Ga}$
does not include corrections on the external legs because it contains
renormalized fields. We shall calculate it in the one-loop
approximation (Fig.~\ref{FigHL}) in the Feynman gauge. We are interested
only in the ultraviolet divergence of the one-loop diagram that does not
depend on external momenta. At zero external momenta we have
\begin{equation}
\Ga\left[1-iC_F g_b^2\int\frac{d^D k}{(2\pi)^D}
\frac{\widehat{k}\ga_0}{(k^2)^2 k_0}\right]
=\Ga\left[1-iC_F g_b^2\int\frac{d^D k}{(2\pi)^D}
\frac1{(k^2)^2}\right]
\label{HL}
\end{equation}
because $\widehat{k}=k_0\ga_0-\vec{k}\cdot\vec{\ga}$ and the integral with
$\vec{k}$ vanishes due to the symmetry. If we started from an infrared
regularized matrix element (e.~g.\ with nonzero external momenta or gluon
mass) we would obtain an integral with the same ultraviolet divergence but
infrared safe. Separating the ultraviolet pole we have
$\Z_\Ga=1+C_F\frac{\al_s}{4\pi\ep}$. Using also~(\ref{ZQ}) and the standard
QCD renormalization constant $Z_q=1-C_F\frac{\al_s}{4\pi\ep}$ in the Feynman
gauge, we obtain the gauge invariant renormalization constant%
~\cite{ShifVol1,PolitzerWise,EichtenHill}
\begin{equation}
\Z_j=1+3C_F\frac{\al_s}{8\pi\ep}.
\label{Zj}
\end{equation}

\begin{figure}[ht]
\begin{center}
\includegraphics{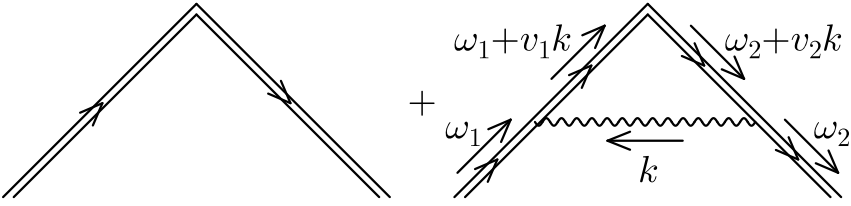}
\end{center}
\caption{Heavy-heavy vertex}
\label{FigHH}
\end{figure}

Similarly, the vertex $\widetilde{\De}=\va{\Q_2|\overline{\Q}_2\Q_1|\Q_1}$
in the one loop approximation (Fig.~\ref{FigHH}) in the Feynman gauge is
\begin{eqnarray}
&&1-iC_F g_b^2\ch\phi\int\frac{d^D k}{(2\pi)^D}
\frac1{k^2(v_1 k+\om_1)(v_2 k+\om_2)}
\label{HH}\\
&=&1-2iC_F g_b^2\ch\phi\int\limits_0^1 dx\int\limits_0^\infty dy
\int\frac{d^D k}{(2\pi)^D}
\nonumber\\
&&\frac1{\left[k^2+y(xv_1+(1-x)v_2)k+y(x\om_1+(1-x)\om_2)\right]^3}
\nonumber\\
&=&1-\frac{C_F g_b^2}{(4\pi)^{D/2}}\Ga(1+\ep)\ch\phi\int\limits_0^1 dx
\int\limits_0^\infty dy y^{-\ep}
\nonumber\\
&&\left[{\textstyle\frac14}(x^2+(1-x)^2+2x(1-x)\ch\phi)y
-x\om_1-(1-x)\om_2\right]^{-1-\ep}
\nonumber\\
&=&1-\frac{C_F g_b^2}{(4\pi)^{D/2}}4\Ga(2\ep)\Ga(1-\ep)\ch\phi\int\limits_0^1
\frac{[-2\om_1x-2\om_2(1-x)]^{-2\ep}dx}{[x^2+(1-x)^2+2x(1-x)\ch\phi]^{1-\ep}}.
\nonumber
\end{eqnarray}
Retaining only the ultraviolet $1/\ep$ pole and using the substitution
$x=\frac12(1+z\cth\frac\phi2)$, we obtain
\begin{equation}
1-C_F\frac{\al_s}{2\pi\ep}\cth\phi\int\limits_{-\th\frac\phi2}^{+\th\frac\phi2}
\frac{dz}{1+z^2},
\label{HH2}
\end{equation}
and finally $Z_\De=1-C_F\frac{\al_s}{2\pi\ep}\phi\cth\phi$. Using~(\ref{ZQ})
we obtain the gauge-invariant renormalization constant $\Z_J=\Z_Q\Z_\De$%
~\cite{Polyakov,Knauss,Korchemsky,FGGW}
\begin{equation}
\Z_J=1-C_F\frac{\al_s}{2\pi\ep}(\phi\cth\phi-1).
\label{ZJ}
\end{equation}
It is equal to 1 at $\phi=0$ because then there is no cusp on the heavy quark
world line.

Knowledge of the renormalization constants $Z$ allows us to determine the
dependence of the renormalized operators on the normalization point $\mu$
using the renormalization group equations. Up to two loops, \MS{}
renormalization constants have the form
\begin{equation}
Z(\mu)=1+\frac{\al_s(\mu)}{4\pi}\frac{c_1}{\ep}
+\left(\frac{\al_s(\mu)}{4\pi}\right)^2
\left(\frac{c_{22}}{\ep^2}+\frac{c_{21}}{\ep}\right)+\cdots
\label{Zform}
\end{equation}
From $g_b=\m^\ep Z_\al^{1/2}g={\rm const}$ we obtain the evolution of
$\al_s(\mu)$: $\frac{d\log\al_s}{d\log\mu}=-2(\ep+\be(\al_s))$,
$\be(\al_s)=\frac12\frac{d\log Z_\al}{d\log\mu}=\be_1\frac{\al_s}{4\pi}
+\be_2\left(\frac{\al_s}{4\pi}\right)^2+\cdots$. It is well known that
$\be_1=\frac{11}{3}N_c-\frac23n_l$,
$\be_2=\frac{34}{3}N_c^2-\left(\frac{13}{3}N_c-\frac1{N_c}\right)n_l$.
Substituting the form~(\ref{Zform}) for $Z_\al$ we see that $c_1=-\be_1$,
$c_{22}=\be_1^2$, $c_{21}=-\frac12\be_2$, i.~e.\ $c_{22}$ is not independent.
Similarly, the $\mu$-dependence of any operator $j$ is usually characterized
by its anomalous dimension $\ga_j=\frac{d\log Z_j}{d\log\mu}
=\ga_1\frac{\al_s}{4\pi}+\ga_2\left(\frac{\al_s}{4\pi}\right)^2+\cdots$.
Substituting the form~(\ref{Zform}) for $Z_j$ we see that $c_1=-\frac12\ga_1$,
$c_{22}=\frac18(\ga_1^2+2\be_1\ga_1)$, $c_{21}=-\frac14\ga_2$, i.~e.\ again
$c_{22}$ is determined by one-loop quantities. In the case of a non-gauge-%
invariant operator, $Z$ also depends on the gauge parameter $a(\mu)$,
and the formulae become more complicated. Solving the renormalization group
equation we obtain
\begin{equation}
j(\mu)=\widehat{\jmath}\,\al_s^{\ga_1/2\be_1}(\mu)
\left[1+\left(\frac{\ga_2}{2\be_1}-\frac{\ga_1\be_2}{2\be_1^2}\right)
\frac{\al_s(\mu)}{4\pi}+\cdots\right],
\label{RenormGroup}
\end{equation}
where $\widehat{\jmath}$ is a renormalization group invariant. This formula
allows us to relate $j(\mu_1)$ to $j(\mu_2)$. Here we present for reference
the two-loop anomalous dimensions of heavy-light and heavy-heavy currents%
~\cite{BroadGr1,Korchemsky}
\begin{eqnarray}
\widetilde{\ga}_j&=&-\frac{3C_F\al_s}{4\pi}
-\left[\frac{49}{96}N_c-\frac{5}{32}C_F-\frac{5}{48}n_l
+(4C_F-N_c)\frac{\pi^2}{24}\right]\frac{C_F\al_s^2}{\pi^2}+\cdots
\nonumber\\
\widetilde{\ga}_J&=&\frac{C_F\al_s}{\pi}(\phi\cth\phi-1)
\label{TwoLoop}\\
&&+\Bigg[-n_l\frac{5}{18}(\phi\cth\phi-1)+N_c\Big(\frac12
+\Big(\frac{67}{36}-\frac{\pi^2}{24}\Big)(\phi\cth\phi-1)
\nonumber\\
&&-\cth\phi\int\limits_0^\phi\psi\cth\psi d\psi
+\cth^2\phi\int\limits_0^\phi\psi(\phi-\psi)\cth\psi d\psi
\nonumber\\
&&-\frac{\sh\phi}{2}\int\limits_0^\phi
\frac{\psi\cth\psi-1}{\sh^2\phi-\sh^2\psi}\log\frac{\sh\phi}{\sh\psi}d\psi
\Big)\Bigg]\frac{C_F\al_s^2}{\pi^2}+\cdots
\nonumber
\end{eqnarray}

Until now we discussed the renormalization inside HQET. But usually we are
interested in matrix elements of QCD operators (e.~g.\ weak currents).
Therefore we have to discuss the relation of QCD operators to their HQET
analogues. Operators in HQET differ from those in QCD starting from the
one-loop level even if written
in the same form via the fields because their matrix elements are calculated
using different Feynman rules. A QCD operator $j$ matches the corresponding
HQET operator $A\j$ if they give identical physical (on-shell) matrix elements
between states suitable for HQET treatment (with residual momenta much less
than $m$). In order to calculate on-shell matrix elements we have to use
the on-shell renormalization scheme in which propagators in the on-shell
limit are free. For the ``massless'' fields $q$, $\Q$ the bare on-shell
propagators get no corrections because loop integrals are no-scale
(ultraviolet and infrared divergences cancel). Therefore the on-shell
renormalized fields coincide with the bare ones: $q=Z_q^{-1/2}q_{\rm os}$,
$\Q=\Z_Q^{-1/2}\Q_{\rm os}$. Note that although the expressions for the
renormalization constants $Z_q$, $\Z_Q$ are the same as above, all
divergences in them are infrared ones because these $Z$ factors relate
renormalized (ultraviolet-finite) fields. For the massive quark field
we have $Q=Z_Q^{-1/2}Q_{\rm os}$,
$Z_Q=1+C_F\frac{\al_s}{4\pi}\left(\frac{2}{\ep}-3L+4\right)$,
where $L=\log\frac{m^2}{\mu^2}$. The infrared divergence of the on-shell
massive quark propagator $Z_Q$ is the same as that of the static quark
propagator $\Z_Q$.

\begin{figure}[ht]
\begin{center}
\includegraphics{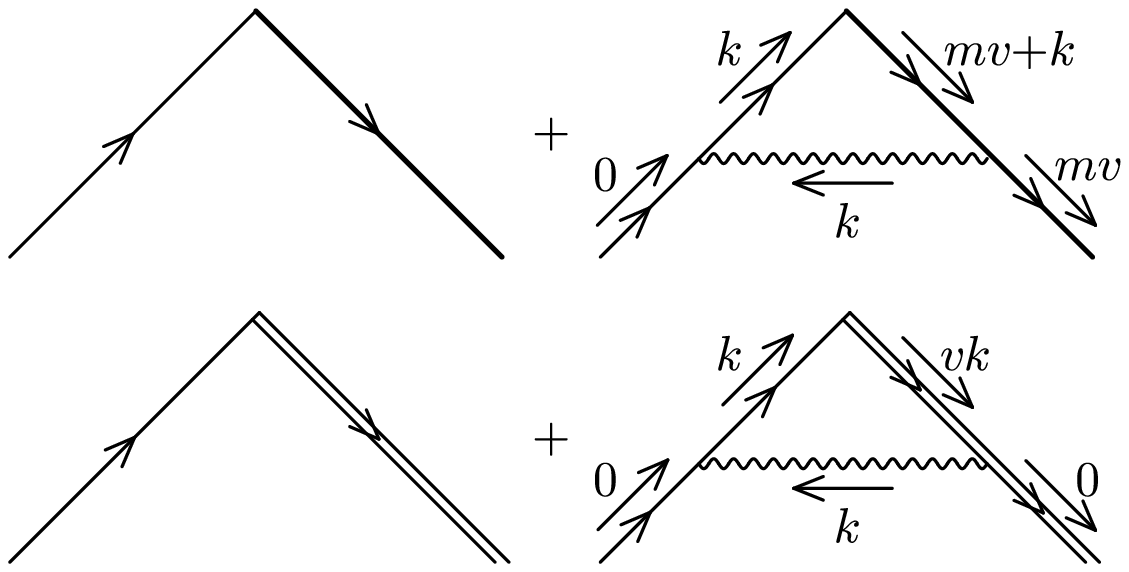}
\end{center}
\caption{Light-light $\to$ heavy-light matching}
\label{FigLLHL}
\end{figure}

\begin{sloppypar}
For the heavy-light bilinear currents we have
$j=Z_\Ga^{-1}Z_q^{-1/2}Z_Q^{-1/2}\overline{Q}_{\rm os}\Ga q_{\rm os}$,
$\j=\Z_\Ga^{-1}Z_q^{-1/2}\Z_Q^{-1/2}\overline{\Q}_{\rm os}\Ga q_{\rm os}$.
Hence the on-shell matrix elements are
$\va{Q|j|q}=Z_\Ga^{-1}Z_q^{-1/2}Z_Q^{-1/2}\Ga$,
$\va{Q|\j|q}=\Z_\Ga^{-1}Z_q^{-1/2}\Z_Q^{-1/2}\widetilde{\Ga}$,
where the proper vertices $\Ga$, $\widetilde{\Ga}$ are depicted on
Fig.~\ref{FigLLHL}. We obtain the matching constant
\begin{equation}
A=A_Q\frac{\Ga/Z_\Ga}{\widetilde{\Ga}/\Z_\Ga},\quad
A_Q=\left(\frac{\Z_Q}{Z_Q}\right)^{1/2}
=1+C_F\frac{\al_s}{4\pi}\left(\frac32L-2\right).
\label{Match}
\end{equation}
Here ultraviolet divergences cancel in $\Ga/Z_\Ga$, $\widetilde{\Ga}/\Z_\Ga$
by definition; infrared divergences cancel between these two expressions
because the infrared behavior of QCD and HQET is identical; $A_Q$ is finite
for the same reason. We choose all quark momenta in HQET to be zero; this
corresponds to the heavy quark momentum $mv$ ($v=(1,\vec0)$) in QCD.
Then the HQET loops (Fig.~\ref{FigLLHL}) vanish: $\widetilde{\Ga}=1$.
Let's calculate the QCD vertex $\Ga$ (Fig.~\ref{FigLLHL}) in the Feynman gauge
\begin{eqnarray}
&&\Ga-iC_F g_b^2\int\frac{d^D k}{(2\pi)^D}
\frac{\ga_\mu(\k+m\v+m)\Ga\k\ga_\mu}{(k^2)^2(k^2+2mvk)}
\label{LLHL1}\\
&=&\Ga-2iC_F g_b^2\int\limits_0^1 dx(1-x)\int\frac{d^D k}{(2\pi)^D}
\nonumber\\
&&\frac{\ga_\mu(\k'+m(1-x)\v+m)\Ga(\k'-mx\v)\ga_\mu}{(k^{\prime2}-m^2 x^2)^3},
\nonumber
\end{eqnarray}
where $k'=k+mxv$. The term with two $\k'$ in the numerator gives
$\frac{k^{\prime2}}{D}H^2(D)\Ga$ where $\ga_\mu\Ga\ga_\mu=H(D)\Ga$;
terms with one $\k'$ vanish. Terms without $\k'$ give
$\pm m^2 x\ga_\mu(1-x+\v)\Ga\ga_\mu=-m^2 x(2\pm H(D)x)\Ga$ where the upper
(lower) sign is for $\Ga$ anticommuting (commuting) with $\v$ and we have
used the fact that $\v$ on the left may be replaced by 1 in the on-shell
matrix element. Therefore all the terms have the common $\ga$-matrix
structure $\Ga$; calculating the integrals, we have the vertex
\begin{eqnarray}
&&\hspace{-20pt}
\Ga=1+C_F\frac{\al_s}{4\pi}\left(\frac{m^2}{\mu^2}\right)^{-\ep}2
\int\limits_0^1dx\,x^{-2\ep}(1-x)
\left[\frac{H^2(D)}{4\ep}+\frac1x\pm\frac{H(D)}2\right]
\nonumber\\&&\hspace{-20pt}
=1+C_F\frac{\al_s}{4\pi}\left(\frac{m^2}{\mu^2}\right)^{-\ep}
\frac1{(1-\ep)(1-2\ep)}
\left[\frac{H^2(D)}{4\ep}-\frac{1-\ep}{\ep}\pm\frac{H(D)}2\right].
\label{LLHL2}
\end{eqnarray}
The first divergence is ultraviolet:
\begin{equation}
Z_\Ga=1+C_F\frac{\al_s}{4\pi}\frac{H^2}{4\ep}.
\label{Zll}
\end{equation}
By the way, the one-loop renormalization constant of the QCD bilinear
quark currents is $Z_j=Z_q Z_\Ga=1+C_F\frac{\al_s}{4\pi}\frac{H^2-4}{4\ep}$;
the vector and axial current ($H=\pm2$) anomalous dimension vanishes.
Finally we obtain from~(\ref{Match}) the matching~\cite{EichtenHill}
\begin{equation}
\overline{Q}\Ga q
=\left[1+C_F\frac{\al_s}{4\pi}\left(-\frac{H^2-10}{4}L+\frac34H^2-HH'
\pm\frac12H-4\right)+\cdots\right]
\overline{\Q}\Ga q,
\label{LLHL}
\end{equation}
where $H'=\frac{dH}{dD}$. This equation holds separately for QCD currents with
$\Ga$ (anti-) commuting with $\ga_0$. If it does not have this property,
it can be split into a commuting and an anticommuting part; it then maps
onto a combination of two HQET currents. The logarithmic part of the matching
constant~(\ref{LLHL}) is determined by the difference of anomalous dimensions
of the QCD and HQET currents; the non-logarithmic part should be included
account only when the two-loop anomalous dimension is also taken into
account~(\ref{RenormGroup}).
\end{sloppypar}

\begin{figure}[ht]
\begin{center}
\includegraphics{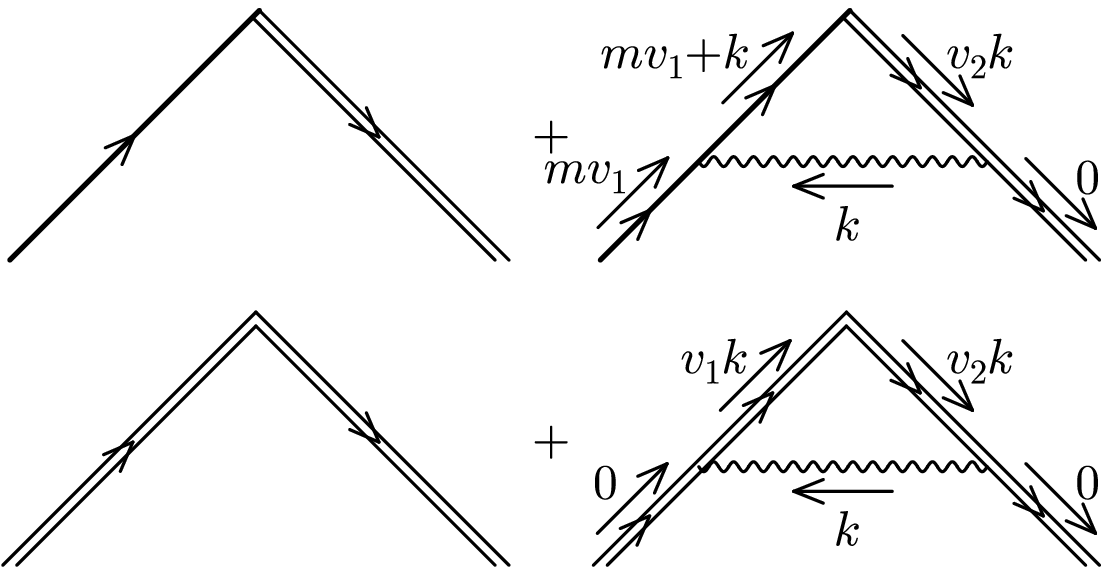}
\end{center}
\caption{Heavy-light $\to$ heavy-heavy matching}
\label{FigHLHH}
\end{figure}

Now we shall consider the current $\overline{\Q}_2\Ga Q_1$ in the effective
theory where $Q_1$ is a heavy quark with the mass $m$. We can go to the
second effective theory in which this quark is also considered static.
The matching is obtained by comparing the diagrams in Fig.~\ref{FigHLHH}.
All external momenta are zero in the second theory, therefore the loops
vanish. In the first theory, the heavy quark has the momentum $mv_1$.
The vertex is the matrix $\Ga$ times
\begin{eqnarray}
&&1-iC_F g_b^2\int\frac{d^D k}{(2\pi)^D}
\frac{(\k+m\v_1+m)\v_2}{k^2(k^2+2mv_1 k)v_2 k}
\label{HLHH1}\\
&=&1-2iC_F g_b^2\int\limits_0^\infty dy\int\limits_0^1 dx
\int\frac{d^D k}{(2\pi)^D}
\frac{-y/2+m(1-x)\v_1\v_2+mv_2}{(k^{\prime2}-y^2/4-m^2 x^2-mxy\ch\phi)^3}
\nonumber
\end{eqnarray}
Now we calculate the loop integrals. The parametric integrals factorize
after using the substitution $y=2mxz$, and the $x$ integrals are trivial.
The upper (lower) sign is for $\Ga$ anticommuting (commuting) with $\v_2$.
\begin{eqnarray}
&&\hspace{-10pt}
1+2C_F\frac{\al_s}{4\pi}\left(\frac{m^2}{\mu^2}\right)^{-\ep}
\int\limits_0^\infty dz\int\limits_0^1 dx\,x^{-1-2\ep}
\frac{xz-2(1-x)\ch\phi\pm x}{(1+z^2+2z\ch\phi)^{1+\ep}}
\label{HLHH2}\\&&\hspace{-10pt}
=1+2C_F\frac{\al_s}{4\pi}\left(\frac{m^2}{\mu^2}\right)^{-\ep}
\int\limits_0^\infty dz
\frac{(z+2\ch\phi\pm1)/(1-2\ep)-\ch\phi/\ep}{(1+z^2+2z\ch\phi)^{1+\ep}}
\nonumber
\end{eqnarray}
The first ultraviolet divergent integral can be calculated by splitting
the integration region at a large $A$ and ignoring $\ep$ in the first
region and $1/z$ in the second one:
\begin{eqnarray}
\int\limits_0^\infty\frac{z\,dz}{(1+z^2+2z\ch\phi)^{1+\ep}}&=&
\int\limits_0^A\frac{z\,dz}{1+z^2+2z\ch\phi}+
\int\limits_A^\infty\frac{dz}{z^{1+2\ep}}
\label{Int1}\\
&=&\log A-\phi\cth\phi+\frac1{2\ep}-\log A.
\nonumber
\end{eqnarray}
The second integral is convergent; we need it up to O($\ep$) because it is
multiplied by the infrared pole $1/\ep$:
\begin{equation}
\int\limits_0^\infty\frac{dz}{(1+z^2+2z\ch\phi)^{1+\ep}}
=\frac1{\sh\phi}\left[\phi-\frac\ep2\left(F(e^{2\phi}-1)-F(e^{-2\phi}-1)
\right)\right],
\label{Int2}
\end{equation}
where
\begin{equation}
F(x)=\int\limits_0^x\frac{\log(1+y)}{y}dy
\label{Spence}
\end{equation}
is the Spence function. Two Spence functions in~(\ref{Int2}) are not
independent: $F(e^{2\phi}-1)+F(e^{-2\phi}-1)=2\phi^2$. The vertex is
\begin{eqnarray}
\Ga&=&1+C_F\frac{\al_s}{4\pi}\Bigg[\frac1\ep-\frac2\ep\phi\cth\phi
-L+2L\phi\cth\phi+2+2\phi\cth\phi\pm2\frac\phi{\sh\phi}
\nonumber\\
&&+\cth\phi\left(F(e^{2\phi}-1)-F(e^{-2\phi}-1)\right)\Bigg].
\label{HLHH3}
\end{eqnarray}
The first divergence is ultraviolet: $Z_\Ga=1+C_F\frac{\al_s}{4\pi}\frac1\ep$;
this is the renormalization constant of the heavy-light current, and it
indeed agrees with what we have found before~(\ref{Zj}). In the denominator
of~(\ref{Match}), we should use the renormalization constant of the
heavy-heavy current $\Z_\De$ found before~(\ref{ZJ}). The infrared divergence
cancels as it should do, and we obtain the matching~\cite{FGGW,Neubert}
\begin{eqnarray}
\overline{\Q}_2\Ga Q_1
&=&\Bigg[1+C_F\frac{\al_s}{4\pi}
\Bigg(2L\phi\cth\phi+\frac12L+2\phi\cth\phi\pm2\frac\phi{\sh\phi}
\nonumber\\
&+&\cth\phi\left(F(e^{2\phi}-1)-F(e^{-2\phi}-1)\right)\Bigg)+\cdots\Bigg]
\overline{\Q}_2\Ga\Q_1
\label{HLHH}
\end{eqnarray}

\begin{figure}[ht]
\begin{center}
\includegraphics[width=\linewidth]{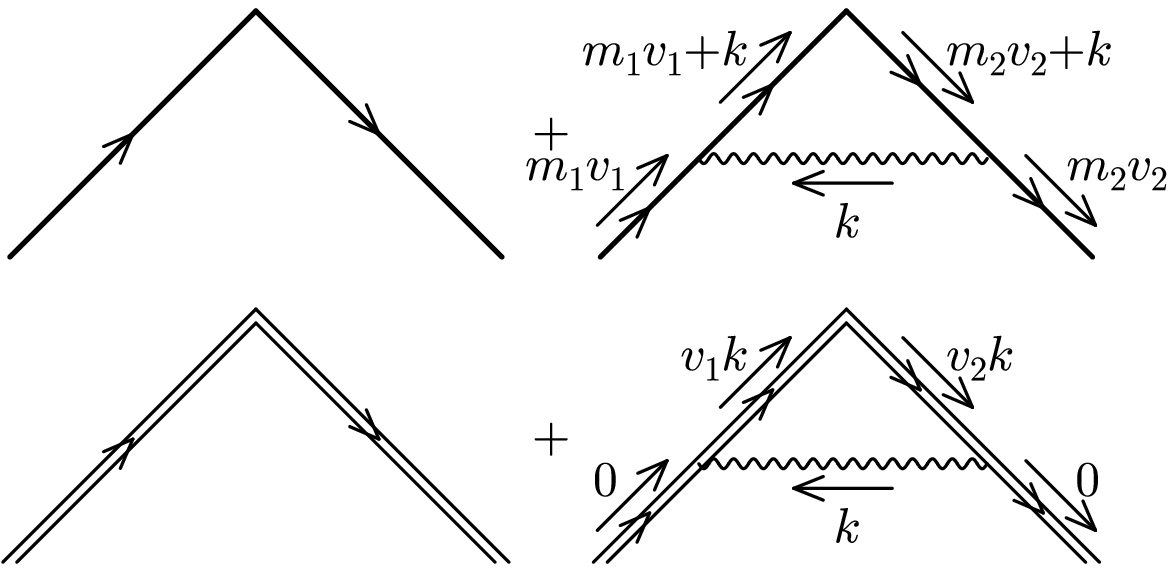}
\end{center}
\caption{Light-light $\to$ heavy-heavy matching}
\label{FigLLHH}
\end{figure}

Finally we shall consider the current $\overline{Q}_2\Ga Q_1$ where $Q_{1,2}$
are the heavy quarks with the masses $m_{1,2}$. We can go to the effective
theory in which both of them are considered static. The matching is obtained
by comparing the diagrams in Fig.~\ref{FigLLHH}. All external momenta are
zero in the effective theory, therefore the loops vanish. In QCD, the heavy
quarks have the momenta $m_{1,2}v_{1,2}$. The vertex is
\begin{eqnarray}
&&\hspace{-20pt}
\Ga-iC_F g_b^2\int\frac{d^D k}{(2\pi)^D}
\frac{\ga_\mu(\k+m_2\v_2+m_2)\Ga(\k+m_1\v_1+m_1)\ga_\mu}%
{k^2(k^2+2m_1 v_1 k)(k^2+2m_2 v_2 k)}
\label{LLHH1}\\
&&\hspace{-20pt}
=\Ga-2iC_F g_b^2\int dx_1dx_2\frac{d^D k}{(2\pi)^D}
\nonumber\\
&&\hspace{-20pt}
\frac{\ga_\mu(\k'+m_2(1-x_2)\v_2-m_1x_1\v_1+m_2)\Ga
(\k'+m_1(1-x_1)\v_1-m_2\v_2+m_1)\ga_\mu}{(k^{\prime2}-a^2)^3},
\nonumber
\end{eqnarray}
where $a^2=m_1^2x_1^2+m_2^2x_2^2+2m_1m_2x_1x_2\ch\phi$.
Now we calculate the loop integrals. The parametric integrals factorize after
the substitution $x_{1,2}=x(1\pm z)/2$, and the $x$ integrals are trivial
($a^2=m_1m_2x^2a_+a_-$,
$a_\pm=\ch\frac{\psi\pm\phi}{2}+z\sh\frac{\psi\pm\phi}{2}$,
where $\psi=\log\frac{m_1}{m_2}$):
\begin{eqnarray}
&&\Ga\Bigg\{1+C_F\frac{\al_s}{4\pi}\left(\frac{m_1m_2}{\mu^2}\right)^{-\ep}
\int\limits_{-1}^{+1}\frac{dz}{(a_+a_-)^\ep}
\Bigg[\frac{H^2(D)}{8\ep(1-\ep)}+\frac{\ch\phi}{\ep(1-2\ep)a_+a_-}
\nonumber\\&&\quad
+\frac{\ch\psi+z\sh\psi}{(1-2\ep)a_+a_-}
-\frac{H(D)(1-z^2)}{16(1-\ep)a_+a_-}\Bigg]\Bigg\}
\label{LLHH2}\\
&&-\v_1\Ga C_F\frac{\al_s}{4\pi}\frac12\int\limits_{-1}^{+1}
\frac{dz}{a_+a_-}\left[1-z+{\textstyle\frac18}He^\psi(1+z)^2\right]
\nonumber\\
&&-\Ga\v_2 C_F\frac{\al_s}{4\pi}\frac12\int\limits_{-1}^{+1}
\frac{dz}{a_+a_-}\left[1+z+{\textstyle\frac18}He^{-\psi}(1-z)^2\right]
\nonumber\\
&&-\v_1\Ga\v_2 C_F\frac{\al_s}{4\pi}\frac{H}{16}\int\frac{1-z^2}{a_+a_-}dz.
\nonumber
\end{eqnarray}
At $m_1=0$ ($\psi\to-\infty$) this vertex reduces to~(\ref{LLHL2}), and at
$m_2=0$ ($\psi\to+\infty$)---to the mirror symmetric expression. In order to
check this, we should take the limit before $\ep\to0$. Ultraviolet
divergences are the same~(\ref{Zll}), but the structure of infrared $1/\ep$
poles is different at zero and nonzero masses. The integrals in $z$ are easily
calculable:
\begin{eqnarray}
&&\int\limits_{-1}^{+1}\frac{dz}{(a_+a_-)^\ep}=2-2\ep\left(
\frac{\phi\sh\phi-\psi\sh\psi}{\ch\phi-\ch\psi}-2\right),
\label{IntLLHH}\\
&&\int\limits_{-1}^{+1}\frac{dz}{(a_+a_-)^{1+\ep}}=\frac1{\sh\phi}
\Bigg\{2\phi+\ep\Bigg[\left(
F\left(\frac{e^\phi-e^{-\phi}}{e^\psi-e^\phi}\right)+(\psi\to-\psi)
\right)
\nonumber\\&&\quad
-(\phi\to-\phi)
+2\psi\log\frac{\sh\frac{\psi+\phi}{2}}{\sh\frac{\psi-\phi}{2}}\Bigg]\Bigg\},
\nonumber\\
&&\int\limits_{-1}^{+1}\frac{z\,dz}{a_+a_-}=\frac{2}{\sh\phi}
\frac{\phi\sh\psi-\psi\sh\phi}{\ch\phi-\ch\psi},
\nonumber\\
&&\int\limits_{-1}^{+1}\frac{z^2 dz}{a_+a_-}=
-\frac{4}{\ch\phi-\ch\psi}-\frac{4\psi\sh\psi}{(\ch\phi-\ch\psi)^2}
+\frac{2\phi}{\sh\phi}\frac{\sh^2\phi+\sh^2\psi}{(\ch\phi-\ch\psi)^2}.
\nonumber
\end{eqnarray}
The Spence functions here are not independent:
$F\left(\frac{e^\phi-e^{-\phi}}{e^{\pm\psi}-e^\phi}\right)+(\phi\to-\phi)
=\frac12\log^2\frac{e^{\pm\psi}-e^\phi}{e^{\pm\psi}-e^{-\phi}}$.
Matching is determined by the formula similar to~(\ref{Match}) but with
$A_{Q1}A_{Q2}$; infrared divergences cancel as they should do, and we finally
obtain~\cite{e+e-2,Neubert}
\begin{eqnarray}
&&\hspace{-20pt}
\overline{Q}_2\Ga Q_1=A\overline{\Q}_2\Ga\Q_1+A_1\overline{\Q}_2\v_1\Ga\Q_1
+A_2\overline{\Q}_2\Ga\v_2\Q_1+A_{12}\overline{\Q}_2\v_1\Ga\v_2\Q_1,
\label{LLHH}\\
&&\hspace{-20pt}
A=1+C_F\frac{\al_s}{4\pi}\Bigg[
-\left(\frac{H^2}{4}+2\phi\cth\phi-3\right)L
-\frac{H^2}{4}\left(\frac{\phi\sh\phi-\psi\sh\psi}{\ch\phi-\ch\psi}-3\right)
\nonumber\\&&\hspace{-20pt}\quad
-HH'+\frac{H}{4}\left(
\frac{\phi}{\sh\phi}\frac{\ch\phi\ch\psi-1}{(\ch\phi-\ch\psi)^2}
-\frac{\psi\sh\psi}{(\ch\phi-\ch\psi)^2}
-\frac1{\ch\phi-\ch\psi}\right)
\nonumber\\&&\hspace{-20pt}\quad
+\cth\phi\left(\left(F\left(\frac{e^\phi-e^{-\phi}}{e^\psi-e^\phi}\right)
+(\psi\to-\psi)\right)-(\phi\to-\phi)
+2\psi\log\frac{\sh\frac{\psi+\phi}{2}}{\sh\frac{\psi-\phi}{2}}\right)
\nonumber\\&&\hspace{-20pt}\quad
-2\frac{\phi}{\sh\phi}\frac{\ch\phi\ch\psi-2\ch^2\phi+1}{\ch\phi-\ch\psi}
-2\frac{\psi\sh\psi}{\ch\phi-\ch\psi}-4\Bigg],
\nonumber\\&&\hspace{-20pt}
A_1=-C_F\frac{\al_s}{4\pi}\Bigg[
\frac14He^\psi\Bigg(\frac{\phi}{\sh\phi}
\left(\frac{\ch\phi\ch\psi-1}{(\ch\phi-\ch\psi)^2}
+\frac{\phi\sh\psi-\psi\sh\phi}{\ch\phi-\ch\psi}+1\right)
\nonumber\\&&\hspace{-20pt}\quad
-\frac{\psi\sh\psi}{(\ch\phi-\ch\psi)^2}-\frac1{\ch\phi-\ch\psi}\Bigg)
+2\frac{\phi}{\sh\phi}\left(1-\frac{\phi\sh\psi-\psi\sh\phi}{\ch\phi-\ch\psi}
\right)\Bigg],
\nonumber\\&&\hspace{-20pt}
A_2=-C_F\frac{\al_s}{4\pi}\Bigg[
\frac14He^{-\psi}\Bigg(\frac{\phi}{\sh\phi}
\left(\frac{\ch\phi\ch\psi-1}{(\ch\phi-\ch\psi)^2}
-\frac{\phi\sh\psi-\psi\sh\phi}{\ch\phi-\ch\psi}+1\right)
\nonumber\\&&\hspace{-20pt}\quad
-\frac{\psi\sh\psi}{(\ch\phi-\ch\psi)^2}-\frac1{\ch\phi-\ch\psi}\Bigg)
+2\frac{\phi}{\sh\phi}\left(1+\frac{\phi\sh\psi-\psi\sh\phi}{\ch\phi-\ch\psi}
\right)\Bigg],
\nonumber\\&&\hspace{-20pt}
A_{12}=C_F\frac{\al_s}{4\pi}\frac{H}{4}\left(
\frac{\phi}{\sh\phi}\frac{\ch\phi\ch\psi-1}{(\ch\phi-\ch\psi)^2}
-\frac{\psi\sh\psi}{(\ch\phi-\ch\psi)^2}-\frac1{\ch\phi-\ch\psi}\right),
\nonumber
\end{eqnarray}
where $L=\log\frac{m_1m_2}{\mu^2}$.

The one-loop matching of the baryonic currents was considered in~\cite{GrYak}.

Now we are in a position to make some statements of Sec.~\ref{SecMeson},
\ref{SecBaryon} more precise. The QCD meson constants~(\ref{Scaling}) are
related to the HQET constant by the matching~(\ref{LLHL}):
\begin{equation}
f=\frac{2\f(m)}{\sqrt{m}}\left(1-c\frac{\al_s(m)}{\pi}+\cdots\right),
\label{Scaling2}
\end{equation}
where $c=\frac43$ for vector mesons and $c=\frac23$ for pseudoscalar mesons
(if a fully anticommuting $\ga_5$ is used). The HQET constant depends on
the normalization point as~(\ref{RenormGroup}--\ref{TwoLoop}):
\begin{equation}
\f(\mu)=\widehat{f}\al_s^{-2/\be_1}(\mu)\left(1-k\frac{\al_s(\mu)}{\pi}
+\cdots\right),\quad
k=\frac{5}{12}-\frac{285-7\pi^2}{27\be_1}+\frac{107}{2\be_1^2}.
\label{Scaling3}
\end{equation}

There are two approaches to the $b\to c$ weak decays: one-step matching%
~\cite{e+e-2,Neubert} and two-step matching~\cite{FGGW,FalkGrin,Neubert}.
We are interested in hadronic matrix elements of the vector and axial
weak currents $j=\overline{c}\Ga b$, $\Ga=\ga_\mu$ or $\ga_\mu\ga_5$. These
currents are defined in QCD at a high normalization point $\mu\sim m_W$.
In the first approach, we use QCD at the scales from $m_W$ down to some
not exactly definable border $\overline{m}\sim m_b\sim m_c$. By a chance,
the QCD anomalous dimensions of these currents vanish, and
$j(\overline{m})=j(m_W)$. At $\mu=\overline{m}$ we perform the matching%
~(\ref{LLHH}) to the HQET in which both $b$ and $c$ quarks are considered
static. The vector current becomes a combination of
$\overline{\widetilde{c}}\Ga\widetilde{b}$ with $\Ga=\ga_\mu$, $v_{b\mu}$,
and $v_{c\mu}$; the axial current---of the similar currents with the extra
$\ga_5$. Then we scale down to a typical hadronic $\mu$ using the HQET
heavy-heavy anomalous dimension~(\ref{TwoLoop}). At this point we use the
heavy quark spin symmetry, and express the matrix elements via the Isgur-Wise
form factors.

In the two-step approach, we use QCD from $\mu=m_W$ down to $\mu=m_b$.
At this point we perform the matching~(\ref{LLHL}) to the HQET-1 in which
$b$ is static while $c$ is still dynamic. The vector current becomes a
combination of $\overline{c}\Ga\widetilde{b}$ with $\Ga=\ga_\mu$ and
$v_{b\mu}$ (and the extra $\ga_5$ in the axial case). Then we use the HQET
heavy-light anomalous dimension~(\ref{TwoLoop}) to scale these currents
down to $\mu=m_c$. At this point we perform the matching~(\ref{HLHH}) to
the HQET-2 in which both $b$ and $c$ are static. We obtain a combination of
$\overline{\widetilde{c}}\Ga\widetilde{b}$ with $\Ga=\ga_\mu$, $v_{b\mu}$,
and $v_{c\mu}$ (with the extra $\ga_5$ in the axial case). Then we use
the HQET heavy-heavy anomalous dimension~(\ref{TwoLoop}) and the spin
symmetry as before.

In the one-step approach, we can't sum the $\al_s\log\frac{m_b}{m_c}$
corrections; we can do it in the two-step approach (even in the subleading
order). On the other hand, the first matching in the two-step method
gives a series in $\frac{m_c}{m_b}$ because $m_c$ is the largest mass scale
in the intermediate HQET. In the above description all $\frac{m_c}{m_b}$
corrections were discarded, and this is not a good approximation in the
real world. The first $\frac{m_c}{m_b}$ correction can be included%
~\cite{FalkGrin} (the leading $\al_s\log\frac{m_b}{m_c}$ corrections are
summed in this term using the one-loop anomalous dimensions), but
incorporating the second term would require a large work.

The one-step matching seems more adequate in our world in which
$\frac{m_c}{m_b}$ is not very small and $\log\frac{m_b}{m_c}$ is not too large.
It is possible to obtain the optimal combination of both approaches%
~\cite{Neubert}. We expand the result of the one-step matching in
$\frac{m_c}{m_b}$. Then we extract the zeroth term from the series, and
replace it by the result of the two-step matching. We also extract the
first term, and replace it by the first power correction from the two-step
matching. The errors of this procedure are of the order of $\al_s^2$,
or $\frac{m_c}{m_b}\al_s$, or
$\left(\frac{m_c}{m_b}\right)^2\al_s\log\frac{m_b}{m_c}$;
they all are small.

\end{document}